\documentclass[aps,pra,reprint,superscriptaddress,showpacs]{revtex4-1}
\usepackage{amsfonts,amsmath,amssymb,amsthm}

\usepackage{array,bm,color,dcolumn,epsfig,graphicx,times,url}
\usepackage[colorlinks=true]{hyperref} 
\hypersetup{colorlinks=true, pdfauthor={Robert L. Kosut et al.},
  pdftitle={Robust control of quantum gates via sequential convex
    programming},
citecolor={blue}, filecolor={blue}, linkcolor={blue}, urlcolor={blue}}

\newtheorem{algorithm}{Algorithm}

\begin{document}

\title{Robust control of quantum gates via sequential convex programming}

\author{Robert L.~Kosut}
\email[Electronic address: ]{kosut@scsolutions.com}
\affiliation{SC Solutions, Inc., 1261 Oakmead Parkway, Sunnyvale, CA
  94085}
\author{Matthew D.~Grace}
\email[Electronic address: ]{mgrace@sandia.gov}
\author{Constantin Brif}
\email[Electronic address: ]{cnbrif@sandia.gov}
\affiliation{Department of Scalable \& Secure Systems Research, Sandia
  National Laboratories, Livermore, CA 94550} 

\date{\today}

\begin{abstract}  
Resource tradeoffs can often be established by solving an appropriate
robust optimization problem for a variety of scenarios involving
constraints on optimization variables and uncertainties. Using an
approach based on sequential convex programming, we demonstrate that
quantum gate transformations can be made substantially robust against
uncertainties while simultaneously using limited resources of control
amplitude and bandwidth. Achieving such a high degree of robustness
requires a quantitative model that specifies the range and character
of the uncertainties. Using a model of a controlled one-qubit system
for illustrative simulations, we identify robust control fields for a
universal gate set and explore the tradeoff between the worst-case
gate fidelity and the field fluence. Our results demonstrate that,
even for this simple model, there exist a rich variety of control
design possibilities. In addition, we study the effect of noise
represented by a stochastic uncertainty model.
\end{abstract}
\pacs{03.67.--a, 02.30.Yy, 02.60.Pn}

\maketitle

\section{Introduction}

Robust control and robust optimization of uncertain systems are
essential in many areas of science and engineering
\cite{Weinmann.book.1991, Zhou.Doyle.1998,
Dullerud.Paganini.book.2000, Lin.book.2007, Belmiloudi.book.2008,
Taguchi.book.2004, Ben-Tal.book.2009, BoydV:04}. Recently, there has
been much interest in achieving robust control of quantum information
systems in the presence of uncertainty
\cite{Fortunato.JCP.116.7599.2002, Pravia.JCP.119.9993.2003,
Boulant.JCP.121.2955.2004, Henry.QIP.6.431.2007,
Borneman.JMR.207.220.2010, Wesenberg.PRA.68.012320.2003,
Wesenberg.PRA.69.042323.2004, Roos.PRA.69.022321.2004,
Skinner.JMR.163.8.2003, Kobzar.JMR.170.236.2004,
Kobzar.JMR.173.229.2005, Luy.JMR.176.179.2005,
Khaneja.JMR.172.296.2005, Skinner.JMR.179.241.2006,
Timoney.PRA.77.052334.2008, Khaneja.JCP.124.114503.2006,
Pryor.Khaneja.JCP.125.194111.2006, Li.Khaneja.PRA.73.030302.2006,
Owrutsky.Khaneja.PRA.86.022315.2012, Ruths.Li.JCP.134.044128.2011,
Ruths.Li.arXiv.1102.3713.2011, Steffen.Koch.PRA.75.062326.2007,
Testolin.PRA.76.012302.2007, Mischuck.PRA.81.023403.2010,
Mischuck.PRA.85.022302.2012, Khani.PRA.85.022306.2012,
Grace.PRA.85.052313.2012, Wang.NatCommun.3.997.2012,
Stihl.arXiv.1210.4311.2012, Green.NJP.15.095004.2013,
QuirozLidar.PRA.88.052306.2013, Brif:QEC11}.
An important property of quantum information processing that
distinguishes it from most other applications is the requirement of an
unprecedented degree of precision in controlling the system
dynamics. Also, due to the very fast time-scale of physical processes
in the quantum realm, implementing closed-loop feedback control is
extremely difficult and thus open-loop control arises as the most
feasible option in most circumstances.

For quantum information systems, a robust optimization problem can be
formulated as a search for \emph{design variables} $\theta \in \Theta$
(where $\Theta$ is the \emph{design set}) that maximize a measure of
\emph{quantum gate fidelity} $\mathcal{F}$ over a range of
\emph{uncertain parameters} $\delta \in \Delta$ (where $\Delta$ is the
\emph{uncertainty set}). Fidelity compares a target unitary
transformation with the actual quantum channel which depends on both
$\theta$ and $\delta$. Fidelity is typically normalized: $\mathcal{F}
\in [0,1]$, and the maximum value $\mathcal{F} = 1$ corresponds to a
perfect generation of the target transformation. The design variables
$\theta$ can include time-dependent control fields (for both open-loop
and closed-loop control), measurement configurations (for closed-loop
feedback control), constants associated with physical implementation,
the circuit lay-out, and so on. The uncertainties $\delta$ can affect
any element of the system Hamiltonian (including the design
variables), with specific manifestations and ranges depending on
details of the physical implementation and external hardware. For
example, uncertainties can represent dispersion and/or slow time
variation of parameters such as coupling strengths, exchange
interactions, and applied electromagnetic fields, as well as additive
and/or multiplicative noise in control fields. The uncertainty set
$\Delta$ can thus, in general, contain deterministic and random
variables. Whatever the case, we assume that $\theta$ and $\delta$ are
constrained to known sets $\Theta$ and $\Delta$, respectively.

One common approach to robust control of quantum gates (e.g., see
Refs.~\cite{Khaneja.JMR.172.296.2005, Brif:QEC11}) is based on
maximizing the \emph{average fidelity} given by 
\begin{equation}\label{eq:favg} 
\mathcal{F}_{\mathrm{avg}}(\theta) = 
\mathbb{E}_{\delta\in\Delta}\{ \mathcal{F}(\theta,\delta) \},
\end{equation} 
where $\mathcal{F}(\theta,\delta)$ denotes the fidelity as a function
of design and uncertain variables, and
$\mathbb{E}_{\delta\in\Delta}\{\cdot\}$ is expectation with respect to
the underlying distribution in $\Delta$. Often the average fidelity is
well approximated as the sum over a discrete sample with associated
probabilities $\{\delta_i \in \Delta, p_i \in [0,1]$\}, i.e.,
$\mathcal{F}_{\mathrm{avg}}(\theta) = \sum_i p_i
\mathcal{F}(\theta,\delta_i)$. While the use of the average fidelity
is applicable in some cases (e.g., when the uncertainty represents
weak random noise), the stringent performance requirements of quantum
information processing make it more appropriate, in general, to
estimate gate errors by using the \emph{worst-case fidelity} with
respect to all uncertainties $\delta \in \Delta$:
\begin{equation}\label{eq:fwc}
\mathcal{F}_{\mathrm{wc}}(\theta)
= \min_{\delta\in\Delta}\mathcal{F}(\theta,\delta).
\end{equation}
Also, \emph{worst-case robust optimization} (or \emph{minimax
optimization}) is a well known approach employed in many classical
problems \cite{Ben-Tal.book.2009, Ben-Tal:1998, Ben-Tal:2002,
ElGhaoui:1998, Bertsimas:2006, Calafiore:2006, Vorobyov:2003,
Lorenz:2005, Rustem.Howe.book.2002, ElGhaoui:2003, Lanckriet:2003,
Zhang:2007, MutapcicB:2009, BertsimasNT:2010a, BertsimasNT:2010b,
MutapcicBFJA:2009, OskooiMNJBJ:2012, ZhangKosut:2013}, and some of the
methods developed for these applications can be adapted for robust
control of quantum gates. The worst-case robust optimization problem
for quantum gate fidelity is formulated as:
\begin{equation}\label{eq:opt} 
\begin{array}{ll} 
\text{maximize}
& \displaystyle \min_{\delta} \mathcal{F}(\theta,\delta) 
\\ 
\text{subject to} 
& \theta\in\Theta,\
\delta\in\Delta .
\end{array} 
\end{equation} 
The goal reflected in the problem \eqref{eq:opt} is to find the design
variables $\theta \in \Theta$ that maximize the worst-case fidelity of
Eq.~\eqref{eq:fwc}.
 
In control applications, the design set $\Theta$ represents the set of
control constraints, and is most often convex or sufficiently well
approximated by a convex set. In some cases so is the uncertainty set
$\Delta$, although this is not necessary for solving~\eqref{eq:opt}.
What makes the problem difficult is that the fidelity is not a convex
function of $\theta$ for any sample $\delta \in \Delta$. Non-convex
optimization problems are common in all of science and engineering and
have engendered numerous numerical approaches to finding local optimal
solutions. In particular, effective methods have been developed in
recent years for worst-case robust optimization with non-convex cost
functions \cite{Zhang:2007, MutapcicB:2009, BertsimasNT:2010a,
BertsimasNT:2010b, MutapcicBFJA:2009, OskooiMNJBJ:2012,
ZhangKosut:2013}.

In optimal control applications, the functional dependence of the
objective (e.g., fidelity) on the control variables is referred to as
the \emph{optimal control landscape}
\cite{Rabitz.Science.303.1998.2004,
Chakrabarti.Rabitz.IRPC.26.671.2007, Brif.NJP.12.075008.2010,
Brif.ACP.148.1.2012}. For an ideal model of a closed quantum system
with no uncertainties, the optimal control landscape for the
generation of unitary transformations has a very favorable topology
\cite{Rabitz.PRA.72.052337.2005, Hsieh.Rabitz.PRA.77.042306.2008,
Ho.PRA.79.013422.2009}. Specifically, provided a number of physically
reasonable conditions are
satisfied~\cite{Pechen.Tannor.PRL.106.120402.2011,
*Rabitz.PRL.108.198901.2012, *Pechen.Tannor.PRL.108.198902.2012,
*Pechen.PRA.86.052117.2012, *Fouquieres.Schirmer.arXiv.1004.3492.2010,
*Pechen.PRA.86.052117.2012}, the landscape is free of local optima,
i.e., there exist one manifold of global minimum solutions (resulting
in $\mathcal{F} = 0$) and one manifold of global maximum solutions
(resulting in $\mathcal{F} = 1$), while all other critical points
reside on saddle-point manifolds~\cite{Rabitz.PRA.72.052337.2005,
Hsieh.Rabitz.PRA.77.042306.2008, Ho.PRA.79.013422.2009}. Such a
favorable landscape topology facilitates easy optimization, as any
gradient-based search (various types of which are popular in quantum
optimal control~\cite{Khaneja.JMR.172.296.2005, Krotov.1996.book,
*Tannor.Kazakov.Orlov.1992.chapter,
*Schirmer.Fouquieres.NJP.13.073029.2011,
*Reich.Ndong.Koch.JCP.136.104103.2012, *Eitan.PRA.83.053426.2011,
Zhu.Rabitz.JCP.109.385.1998, *Maday.Turinici.JCP.118.8191.2003,
*Ohtsuki.JCP.120.5509.2004, Ndong.PRA.87.043416.2013,
Hohenester.PRB.74.161307.2006, *Grace.JPB.40.S103.2007,
*Grace.JMO.54.2339.2007, *Montangero.PRL.99.170501.2007,
*Wenin.Potz.PRA.78.012358.2008, *Wenin.Potz.PRB.78.165118.2008,
*Roloff.JCTN.6.1837.2009, *Schirmer.PRA.80.030301.2009,
*Rebentrost.PRL.102.090401.2009, *Schulte.JPB.44.154013.2011,
*Floether.NJP.14.073023.2012, Rothman.JCP.123.134104.2005,
*Rothman.PRA.73.053401.2006, *Dominy.Rabitz.JPA.41.205305.2008,
*Moore.Rabitz.PRA.84.012109.2011, Caneva.PRA.84.022326.2011,
Fouquieres.JMR.212.412.2011, *Machnes.PRA.84.022305.2011,
vonWinckel.Borzi.CPC.181.2158.2010, *vonWinckel.SIAM-JSC.31.4176.2010,
Fouquieres.PRL.108.110504.2012, Degani.Zanna.SIAM-JSC.34.A1488.2012})
is guaranteed to reach the global maximum
\cite{Moore.Chakrabarti.PRA.83.012326.2011}. Unfortunately, when
uncertainties are present, this landscape topology is not
preserved. Typically, uncertainties cause a decrease and fragmentation
of the global maximum manifold, resulting in the emergence of multiple
local maxima \cite{Brif:QEC11} (the landscape also undergoes a similar
transformation when control fields are severely constrained
\cite{Moore.Rabitz.JCP.137.134113.2012}). Provided that the range of
uncertainty is not too large, many of these local optimal solutions
will have fidelities close to one.

For quantum information systems, there is considerable on-going effort
to develop efficient methods for obtaining a good solution to the
problem of robust control, for either average or worst-case
fidelity. The majority of existing approaches rely on a numerical
optimization procedure, mostly involving a gradient-based search for
maximizing the average fidelity of Eq.~\eqref{eq:favg}. In some cases,
a randomized search such as a genetic algorithm is employed
\cite{Brif:QEC11}. The results demonstrate the existence of many
solutions with high fidelities, consistent with the control landscape
picture discussed above. Additionally, the optimal controls are often
similar to the corresponding initial controls, provided the latter are
reasonably good. This phenomenon, also observed in many engineering
and design applications employing local search algorithms, supports
the need for developing tools to efficiently calculate a good initial
control. In particular, empirical evidence and simulations suggest
that robust controls for an uncertain quantum system can be found by
searches that start from solutions generated by applying \emph{optimal
control theory} or \emph{dynamical decoupling} to the ideal
(zero-uncertainty) counterpart system (see, e.g.,
\cite{Brif.ACP.148.1.2012,QuirozLidar.PRA.88.052306.2013} and references
therein).

In this paper, we propose the use of \emph{sequential convex
programming} (SCP) which is one of several methods available for
numerically solving optimization problems like \eqref{eq:opt}. (See
\cite{SCP.1995} for a collection of earlier SCP varieties and uses and
\cite{ee364b} for a recent informative overview.) SCP provides a
general framework for finding local optimal solutions to the
worst-case robust optimization problem \eqref{eq:opt}. The specific
SCP algorithm used here, delineated in Algorithm~\ref{alg:scp} below,
follows directly from \cite{MutapcicB:2009, MutapcicBFJA:2009}. It was
used previously for robust design of slow-light tapers in
photonic-crystal waveguides \cite{MutapcicBFJA:2009, OskooiMNJBJ:2012}
and quantum potential profiles for electron transmission in
semiconductor nanodevices \cite{ZhangKosut:2013}. In this paper, we
apply this SCP algorithm to identify robust control fields for the
generation of quantum gates in an uncertain one-qubit system.

\section{Sequential Convex Programming}

The SCP algorithm used here is shown in abstract form in
Algorithm~\ref{alg:scp}. The algorithm is initialized with (i) a
control in the feasible set $\Theta$, which is assumed to be convex,
(ii) samples $\delta_i,\ i=1,\ldots,L$ taken from the uncertainty set
$\Delta$, which need not be convex, and (iii) a convex trust region
$\tilde{\Theta}_{\mathrm{trust}}$. The trust region is selected so
that the linearized fidelity $\mathcal{F}(\theta,\delta_i) +
\tilde{\theta}^{\mathsf{T}} \nabla_{\theta}
\mathcal{F}(\theta,\delta_i)$, where $\tilde{\theta} \in
\tilde{\Theta}_{\mathrm{trust}}$, used in the optimization step
retains sufficient accuracy. In each iteration the SCP algorithm
returns the optimal increment $\tilde{\theta}$ and the associated
worst-case linearized fidelity. To compute the actual worst-case
fidelity requires simulating the system's evolution with the control
variables $\theta+\tilde{\theta}$ as indicated in Step 3 of
Algorithm~\ref{alg:scp}. The centerpiece is the optimization step
which, in the version shown in Algorithm~\ref{alg:scp}, is gradient
based, thereby resulting in $L$ affine constraints in
$\tilde{\theta}$, and hence is a convex optimization. The Hessian,
perhaps not so easily computed, can be easily incorporated as shown in
Appendix~\ref{sec:scpcode}. In some cases the number of samples, $L$,
can be very large. Fortunately, however, computational complexity
grows gracefully with the number of constraints and thus does not
grossly affect the convex optimization efficiency \cite{BoydV:04}.

\vspace{1ex}\noindent
\hrulefill
\vspace{-1ex}
{\small
\begin{algorithm}
{\rm Robust control via SCP.
\label{alg:scp}
\begin{description}
\setlength{\itemsep}{0mm}
\item[Initialize] \textrm{}\\
control $\theta\in\Theta\subseteq\mathbb{R}^N$; \\
uncertainty/noise sample
$\{\delta_i\in\Delta,\ i=1,\ldots,L\}$; \\
trust region
$\tilde{\Theta}_{\mathrm{trust}} \subseteq \mathbb{R}^N$.
\item[Repeat] \textrm{}\vspace*{-2mm}
\begin{enumerate}
\setlength{\itemsep}{1mm}
\item Calculate fidelities and gradients:\vspace*{2mm}\\
$\mathcal{F}(\theta,\delta_i),\ 
\nabla_{\theta} \mathcal{F}(\theta,\delta_i)\in\mathbb{R}^N,\ 
i=1,\ldots,L$.
\item Using the linearized fidelity, solve for the increment
$\tilde{\theta}$ from the convex optimization:\vspace*{2mm}\\
maximize\ \ $\displaystyle \min_i \big[ 
\mathcal{F}(\theta,\delta_i) + \tilde{\theta}^{\mathsf{T}} 
\nabla_{\theta} \mathcal{F}(\theta,\delta_i) \big]$ \\
subject to\ \ \ $\theta+\tilde{\theta}\in\Theta,\ 
\tilde{\theta} \in \tilde{\Theta}_{\mathrm{trust}}$.
\item Update:\vspace*{1mm}\\
\textbf{if}\ \ $\displaystyle 
\min_i \mathcal{F}(\theta+\tilde{\theta},\delta_i) 
> \min_i \mathcal{F}(\theta,\delta_i)$\ \ \textbf{then}
\\
\hspace*{3.75mm}replace $\theta$ by $\theta+\tilde{\theta}$
and increase $\tilde{\Theta}_{\mathrm{trust}}$\\
\textbf{else}\\
\hspace*{3.75mm}decrease $\tilde{\Theta}_{\mathrm{trust}}$\\
\textbf{endif}
\end{enumerate}
\item[Until] Stopping criteria satisfied.
\end{description}
}
\end{algorithm}
}
\vspace{-2ex}\noindent
\hrulefill
\vspace{2ex}

In Appendix~\ref{sec:scpcode} we show how the gradient and Hessian can
be cast in standard forms compatible with freely available software
specifically designed to solve such convex optimization problems. In
general, solving the convex optimization is not the most time
consuming step in the SCP algorithm. The time-burden in each iteration
falls more often on simulations required to compute the fidelities and
gradients (and the Hessian if used) at each uncertainty sample. Of
course, as is the case with numerical simulations of any quantum
information system, there always lurks the exponential scaling with
the number of qubits.

Despite many advantages, SCP is a local optimization method. As such,
there is no way to verify that a globally optimal solution has been
found. Since the fidelity by construction cannot exceed one, it would
seem that at least the maximum is known, so if $\mathcal{F} = 1$ is
achieved, it is an optimal solution.  However, even as we often obtain
fidelities that are extremely close to one, for example,
$\log_{10}(1-\mathcal{F}) \in [-6,-4]$, this does not guarantee that
the algorithm did not miss a better solution. Although a fidelity
value with 4 to 6 nines following the decimal point is effectively one
for most engineering problems, for quantum computing every additional
improvement in fidelity is important, since it can greatly decrease
the physical resources required for fault-tolerant operation.

\section{Sequential Convex Programming for an uncertain qubit}

In this section, we show how to use SCP for robust control of quantum
gates in the presence of common types of uncertainties and
constraints. We consider a one-qubit system modeled by the
time-dependent Hamiltonian ($\hbar = 1$):
\begin{equation}\label{eq:ham1q}
H(t) = c(t) \omega_x X + \omega_z Z, 
\end{equation}
where $c(t)$ is the external control field (a real-valued function of
time defined on the interval $[0, T]$), and $X$ and $Z$ are the
respective Pauli matrices. The real parameters $\omega_x$ and
$\omega_z$ are constant but uncertain over the time interval
$[0,T]$. Correspondingly, the uncertain parameters $\delta$ in
\eqref{eq:opt} are specified by the parameter vector $\omega =
[\omega_x, \omega_z]^{\mathsf{T}}$. 

\subsection{Control generation and constraints}  

The control field $c(t)$ is typically the output of a signal
generation device whose dynamics impose constraints on magnitudes,
bandwidth, and so on. To illustrate the use of SCP we make the
simplifying assumption that the control is piecewise-constant over $N$
uniform time intervals of width $h = T/N$:
\begin{equation}\label{eq:cpwc} 
c(t,\theta) = \theta_{k}\ \ \text{for}\ \ t \in (t_{k-1}, t_k ],\ \
k = 1, \ldots, N,
\end{equation}
where $t_k = k h$. Correspondingly, the design variables $\theta$ in
the optimization problem \eqref{eq:opt} are specified by the vector of
field values $\theta = [\theta_1, \ldots, \theta_N]^{\mathsf{T}}$. The
set $\Theta$ reflects control constraints, typical examples of which
are shown in Table~\ref{tab:thfeas}. The appearance of control
constraints due to signal generation dynamics is discussed in
Appendix~\ref{sec:signal}. 

A couple of important characteristics of the control field, used in
Table~\ref{tab:thfeas}, are the fluence (a measure of the field
energy):
\begin{equation}\label{eq:fluence}
\Phi(\theta) = \int_0^T c^2(t,\theta) d t = \| \theta \|_2^2 h
\end{equation}
and the area (a measure of the field strength):
\begin{equation}\label{eq:area}
A(\theta) = \int_0^T |c(t,\theta)| d t = \| \theta \|_1 h ,
\end{equation} 
where $\| \theta \|_p = \big( \sum_{k=1}^N |\theta_k|^p \big)^{1/p}$
is the vector $L^p$-norm.

\begin{table}[htbp]
\caption{\label{tab:thfeas}Typical control constraints. The bounding
parameters $\alpha$, $\beta$, $\gamma$ are positive constants and
$c^{\min}$, $c^{\max}$ are real constants. Also, $a$ is a real $N
\times N$ matrix and $b$ is a real vector of length $N$.}
\begin{ruledtabular}
\begin{tabular}{ll}
Constraint & Set $\Theta$ \\ \hline
none & $\mathbb{R}^N$ \\
fluence & $\Phi(\theta) \leq \gamma$ \\
magnitude & $c^{\min} \leq c(t,\theta) \leq c^{\max}, \ t \in [0,T]$ \\
slew rate & $|\dot{c}(t,\theta)| \leq \beta, \ t \in [0,T]$ \\
area & $A(\theta) \leq \alpha$ \\
linear & $a \theta = b$ \\
\end{tabular}
\end{ruledtabular}
\end{table}

The list of control constraints in Table~\ref{tab:thfeas} is certainly
not exhaustive. However, since $c(t,\theta)$ is a linear function of
$\theta$, each of these constraints or any combination thereof forms a
convex set in $\mathbb{R}^N$. The bounding parameters in
Table~\ref{tab:thfeas} can also be used as design variables to
establish control resource tradeoffs via SCP. In particular, the
tradeoff between the gate fidelity and the field fluence is explored
in Sec~\ref{sec:tradeoff}.

\subsection{Evolution operator and fidelity} 

For a given realization of the Hamiltonian \eqref{eq:ham1q} (i.e., for
given values of $\omega_x$ and $\omega_z$), the system undergoes a
unitary evolution, governed by the Schr\"{o}dinger equation:
\begin{equation}\label{eq:schro}
i \frac{d}{dt} U(t) = H(t) U(t) , \ \ \ U(0) = I ,
\end{equation}  
where $U(t) \equiv U(t,0)$ is the time-evolution operator (propagator)
from time $t = 0$ to $t$, and $I$ is the identity operator. For the
piecewise-constant control \eqref{eq:cpwc}, the evolution operator
$U(t_k)$ is given by a product of incremental propagators:
\begin{align}
& U(t_k) = U(t_k,t_{k-1}) \cdots U(t_2,t_1) U(t_1,t_0) , 
\label{eq:evol-oper} \\
& U(t_k,t_{k-1}) = \exp \left[ -i h (\theta_{k} \omega_x X 
+ \omega_z Z) \right].
\label{eq:evol-prop-one-step}
\end{align}
In particular, the evolution operator attained at the final time $T$
is $U_T \equiv U(T) = U(t_N)$. This evolution operator is a function
of $\theta$ and $\omega$.

The fidelity of a quantum gate is a measure of alignment between the
target unitary transformation $W$ and the actual final-time evolution
operator $U_T$. Specifically, for the one-qubit system, we use the
fidelity defined as
\begin{equation}\label{eq:fid}
\mathcal{F}(\theta, \omega)
= \frac{1}{4} \left| \mathrm{Tr} \left( W^{\dag} U_T 
\right) \right|^2 .
\end{equation}
This fidelity, normalized to $[0,1]$, is independent of the phase of
either $W$ or $U_T$. Along with fidelity, we will also use the
normalized distance between $W$ and $U_T$, which is defined as
\begin{equation}\label{eq:dist}
\mathcal{D}(\theta, \omega) = 1 - \mathcal{F}(\theta, \omega) .
\end{equation}
In accordance with Eq.~\eqref{eq:dist},
$\mathcal{D}_{\mathrm{avg}}(\theta) = 1 -
\mathcal{F}_{\mathrm{avg}}(\theta)$ and
$\mathcal{D}_{\mathrm{wc}}(\theta) = 1 -
\mathcal{F}_{\mathrm{wc}}(\theta)$.

\subsection{Uncertainty modeling}

One general approach to modeling the uncertainty in the Hamiltonian
parameters $\omega$ is via a \emph{deterministic} (or set-membership)
model:
\begin{equation}\label{eq:deldet}
\Delta = \left\{ \| \Omega^{-1}(\omega - \bar{\omega}) \|_p 
\leq 1 \right\} ,
\end{equation}
where $\bar{\omega} = [\bar{\omega}_x, \bar{\omega}_z]^{\mathsf{T}}$
is the vector of nominal values, $\Omega$ is a positive-definite
weighting matrix (here $2 \times 2$), and $p$ is typically $2$ or
$\infty$. If $p = \infty$ and $\Omega$ is diagonal, then $\omega_x$
and $\omega_z$ are not correlated, in which case Eq.~\eqref{eq:deldet}
reduces to
\begin{equation}\label{eq:deldet1}
\Delta = \left\{
|\omega_x-\bar{\omega}_x|\leq\tilde{\omega}_x,\
|\omega_z-\bar{\omega}_z|\leq\tilde{\omega}_z
\right\} ,
\end{equation}
where $\Omega = \mathrm{diag}(\tilde{\omega}_x,\tilde{\omega}_z)$. If
$\Omega$ is not diagonal, then $\omega_x$ and $\omega_z$ are
correlated, possibly arising, respectively, from an approximation of a
joint Gaussian or uniform distribution, with $\Omega$, typically,
being the covariance matrix associated with a specified confidence
region for the parameters.

The uncertainty in the parameters can often be best described via a
\emph{probabilistic} model, for example,
\begin{equation}\label{eq:delprob}
\Delta = \left\{\mathbb{E}\{\omega\} = \bar{\omega},\
\mathbb{E}\{(\omega-\bar{\omega})(\omega-\bar{\omega})^{\mathsf{T}}\}
= C \right\},
\end{equation}
where $\mathbb{E}\{\cdot\}$ is expectation with respect to the
underlying probability distribution of $\omega$. If this distribution
is Gaussian, then $\Delta = \{\omega \in
\mathcal{N}(\bar{\omega},C)\}$.

Random uncertainty also arises from noise in the control field and/or
environment, best represented by a \emph{stochastic} model. In this
case, the uncertainty set $\Delta$ can have the same form as in
Eq.~\eqref{eq:delprob}, but here the elements of $\omega(t) =
[\omega_x(t),\omega_z(t)]^{\mathsf{T}}$ are stochastic variables with
the moments $\mathbb{E}\{\omega(t)\} = \bar{\omega}$ and
$\mathbb{E}\{[\omega(t)-\bar{\omega}]
[\omega(t')-\bar{\omega}]^{\mathsf{T}}\} = C(t,t')$.

As mentioned above, the uncertainty set $\Delta$ for SCP need not be
convex; for example, the set of Eq.~\eqref{eq:deldet} is convex, but
that of Eq.~\eqref{eq:delprob} is not. Step 2 in
Algorithm~\ref{alg:scp} only requires that the uncertain parameters be
\emph{sampled} from the set $\Delta$. In a numerical example studied
below, we use a simple uniform sampling from an uncertainty set of the
form~\eqref{eq:deldet1}. More sophisticated methods cycle through a
sampling in the optimization step followed by validation on a
different sampled set; bad parameters revealed in the validation step
can be used in a new sampling for a repeat of the optimization step
(e.g., see Ref.~\cite{MutapcicB:2009}).

\subsection{Robust optimization}

Now we can formulate a specific instance of the optimization problem
\eqref{eq:opt}, corresponding to finding a robust control field for
generating a target quantum gate in an uncertain one-qubit
system. Specifically, the goal is to solve for the field values
$\theta \in \mathbb{R}^N$ from the optimization problem:
\begin{equation}\label{eq:optxz}
\begin{array}{ll}
\text{maximize} 
& 
\displaystyle \min_{\omega \in \Delta} \mathcal{F}(\theta, \omega)
\\
\text{subject to}
&
U_T\ \text{obtained from Eq.~\eqref{eq:evol-oper}},
\\
&
\theta \in \Theta\ \text{from a combination of sets in 
Table~\ref{tab:thfeas}},
\\
&
\omega \in \Delta\ \text{from Eq.~\eqref{eq:deldet} or 
Eq.~\eqref{eq:delprob}.} 
\end{array}
\end{equation}
Since $\Theta$ is a convex set and samples are taken from $\Delta$ to
compute gradients of $\mathcal{F}$ with respect to $\theta$, then Step
2 of Algorithm~\ref{alg:scp} will be a convex optimization. 

\section{Robust one-qubit gates}

We use the SCP routine to find robust control fields corresponding to
the following target unitary transformations:
\begin{equation}\label{eq:1qu gates}
W_{\mathrm{I}} = \begin{bmatrix}
1 & 0 \\ 
0 & 1 
\end{bmatrix}, \
W_{\mathrm{H}} = \frac{1}{\sqrt{2}} \begin{bmatrix}
1 & 1 \\ 
1 & -1  
\end{bmatrix}, \
W_{\mathrm{P}} = \begin{bmatrix}
1 & 0 \\ 
0 & e^{i \pi/4}  
\end{bmatrix}.
\end{equation}
Here, $W_{\mathrm{I}}$, $W_{\mathrm{H}}$, and $W_{\mathrm{P}}$
represent the identity, Hadamard, and phase ($\pi/8$) gates,
respectively. Note that $W_{\mathrm{H}}$ and $W_{\mathrm{P}}$ comprise
a universal gate set for one-qubit operations. 

The uncertainty set used for all optimizations presented in this
section is
\begin{equation}\label{eq:delxz}
\Delta = \left\{ 
|\omega_x-1|\leq 0.01,\
|\omega_z-2|\leq 0.20
\right\},
\end{equation}
corresponding to a deterministic model with 1\% control amplitude
uncertainty and 10\% drift magnitude uncertainty. For each target
gate, SCP is used to solve for $\theta\in\mathbb{R}^N$ from:
\begin{equation}\label{eq:ex optxz}
\begin{array}{ll}
\text{maximize} 
& 
\displaystyle 
\min_{\omega \in \Delta} \left\{ \mathcal{F}(\theta, \omega)
= {\textstyle \frac{1}{4}} \left| 
\mathrm{Tr} \left( W^{\dag} U_{T} \right) \right|^{2} \right\}
\\
\text{subject to}
&
U_{T}\ \text{obtained from Eq.~\eqref{eq:evol-oper}},
\\
&
\theta \in \mathbb{R}^N\ \text{(unconstrained)},
\\
&
\omega\in\Delta\ \text{from Eq.~\eqref{eq:delxz}}.
\end{array}
\end{equation}
We obtain solutions of \eqref{eq:ex optxz} for all combinations of $W
\in \{W_{\mathrm{I}},W_{\mathrm{H}},W_{\mathrm{P}}\}$ and $T \in \{ 1,
2, 4 \}$, along with selected values of $N \in \{ 5, 10, 20, 80
\}$. For each SCP optimization presented in this section, we first
used a gradient-based search
\cite{Moore.Chakrabarti.PRA.83.012326.2011, Moore.PRA.86.062309.2012}
to find a control field vector $\theta^{(0)}$ that achieves
$\mathcal{F} \left( \theta^{(0)}, \bar{\omega} \right) \simeq 0.999$
for the nominal parameter values $\bar{\omega}_{x} = 1$ and
$\bar{\omega}_{z} = 2$. For a fixed $\omega$ (i.e., in the absence of
uncertainty), it is easy to achieve unit fidelity to a desired
numerical accuracy, so $\theta^{(0)}$ is a solution which is close to
the top of the landscape, but not fully optimal. Then, $\theta^{(0)}$
was used as the initial field to start the SCP search for the
uncertain system.

\begin{figure*}
\begin{tabular}{l}\textbf{(a)}
\vspace*{-1mm} \\ \vspace*{-1mm}
\begin{tabular}{lr}
\includegraphics[width=83mm]{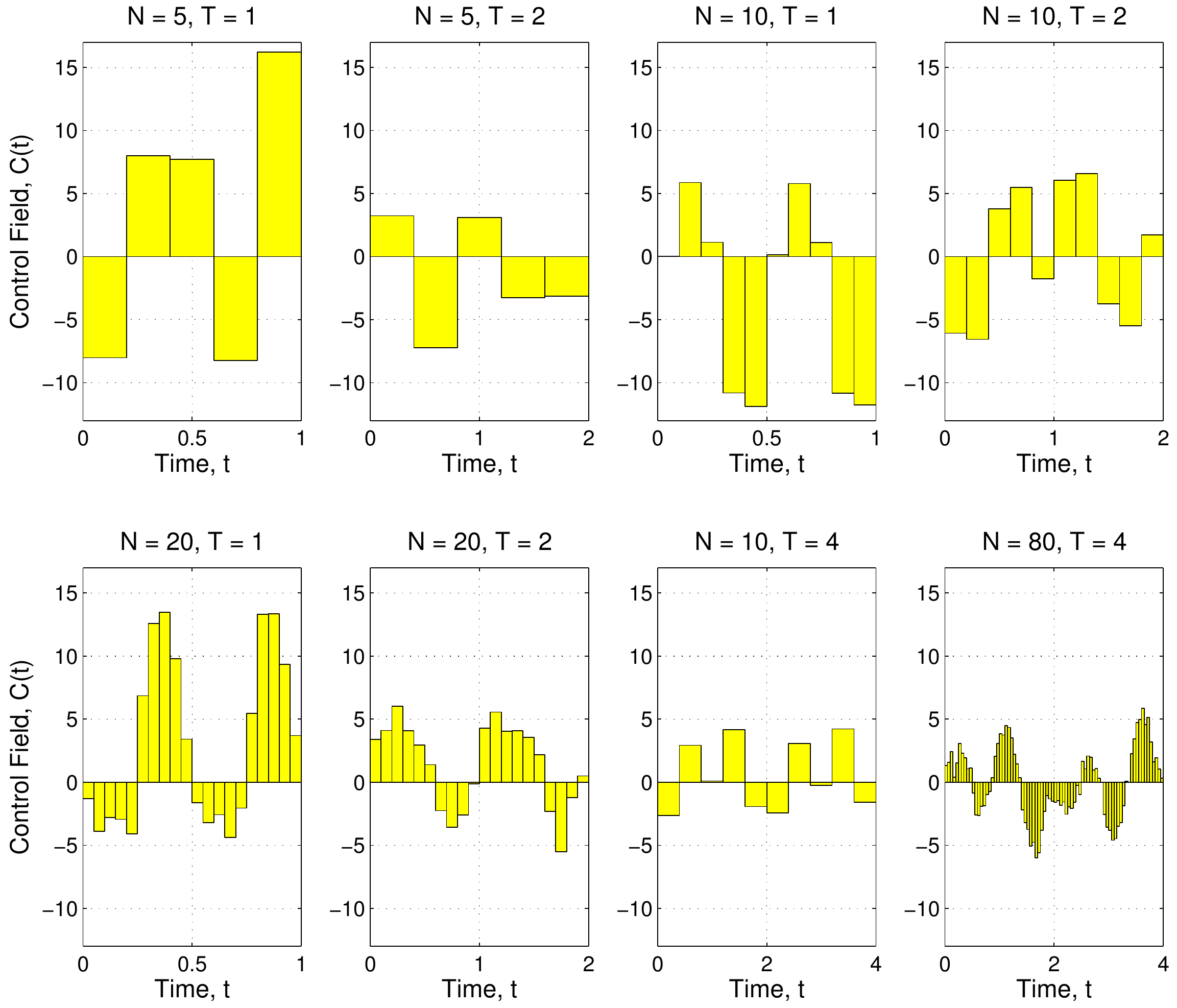}
& \hspace*{5mm}
\includegraphics[width=86.2mm]{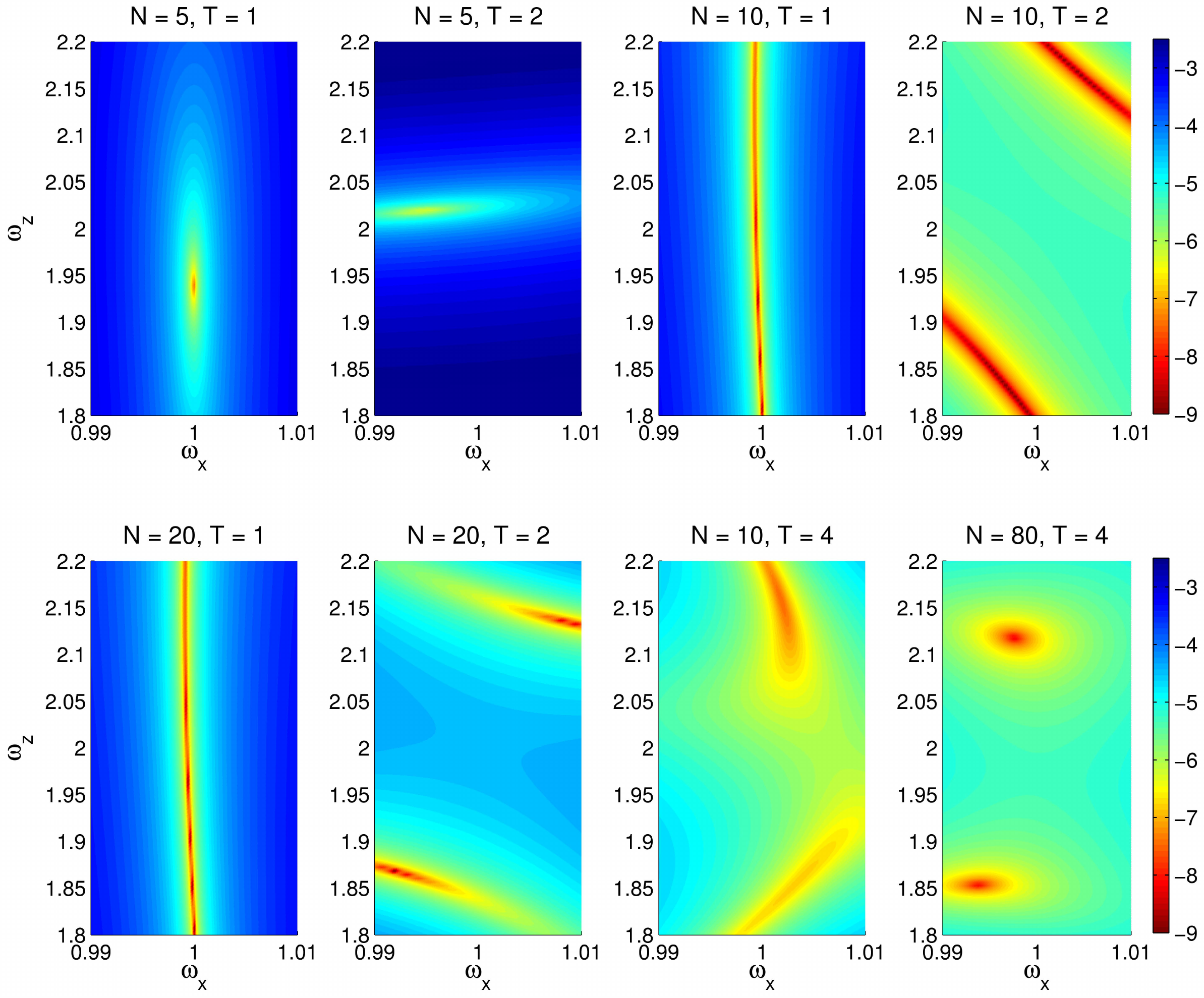}
\end{tabular}
\\
\textbf{(b)}
\vspace*{-1mm} \\ \vspace*{-1mm}
\begin{tabular}{lr}
\includegraphics[width=83mm]{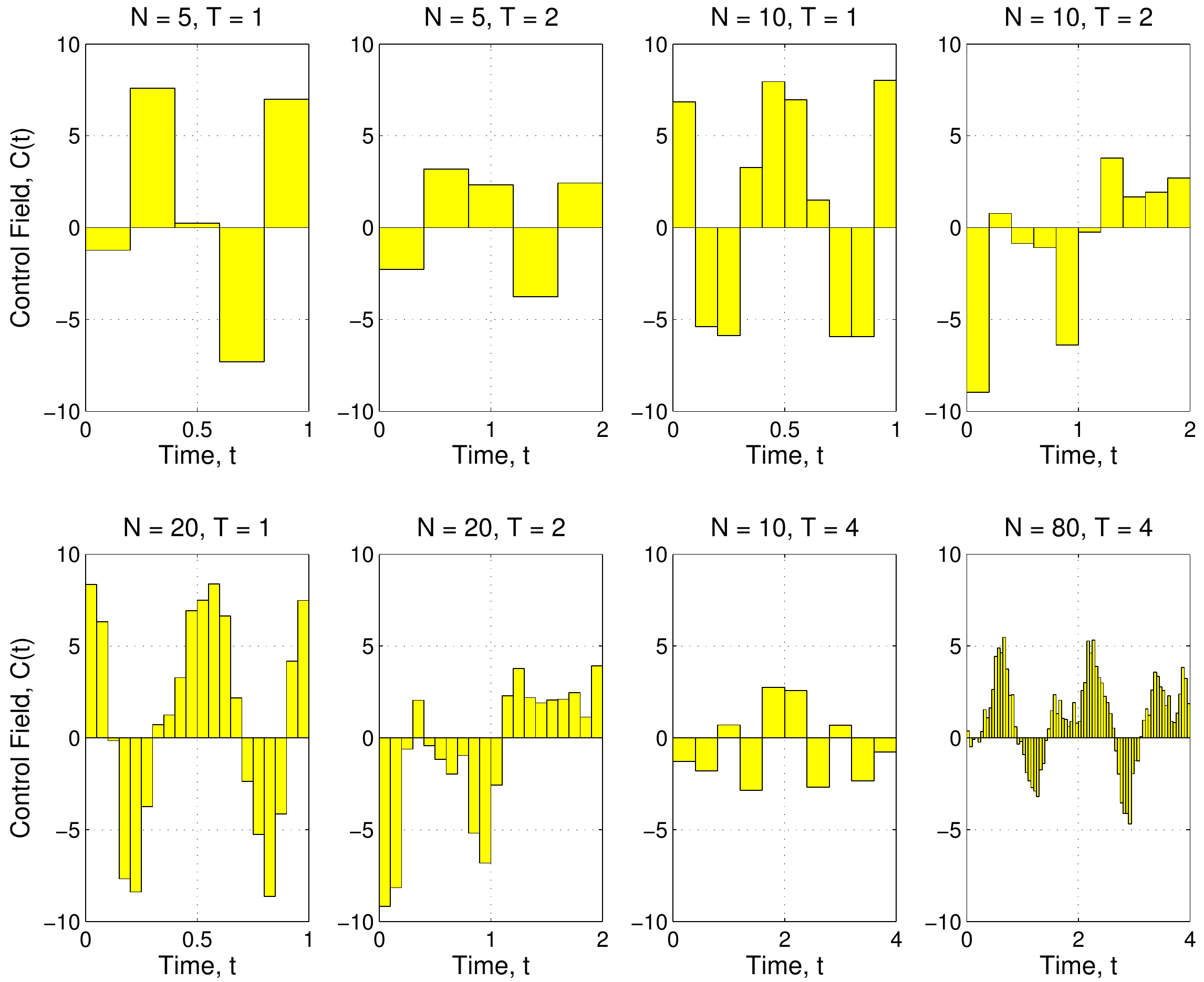}
& \hspace*{5mm}
\includegraphics[width=86.2mm]{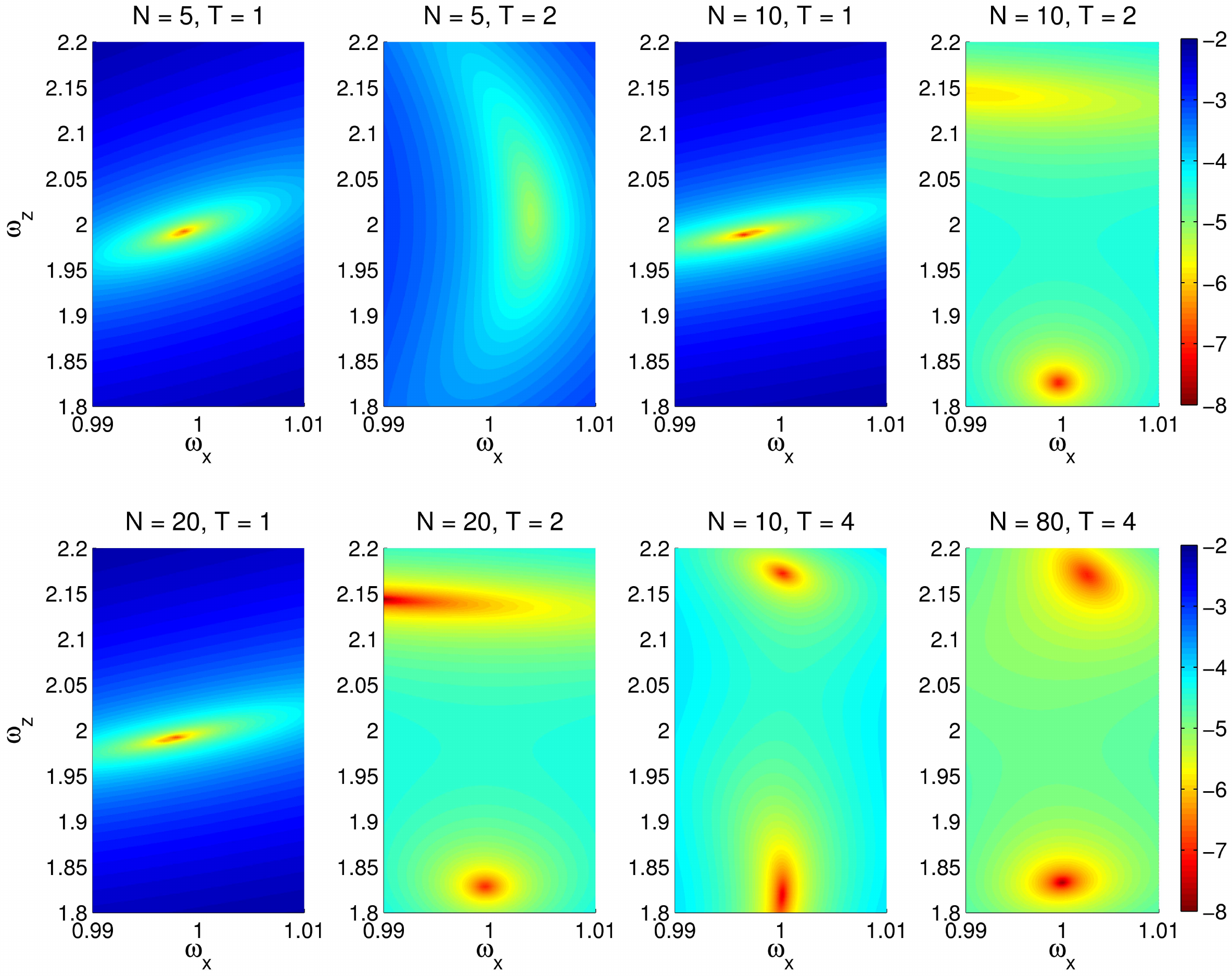}
\end{tabular}
\\
\textbf{(c)}
\vspace*{-1mm} \\ \vspace*{-1mm}
\begin{tabular}{lr}
\includegraphics[width=83mm]{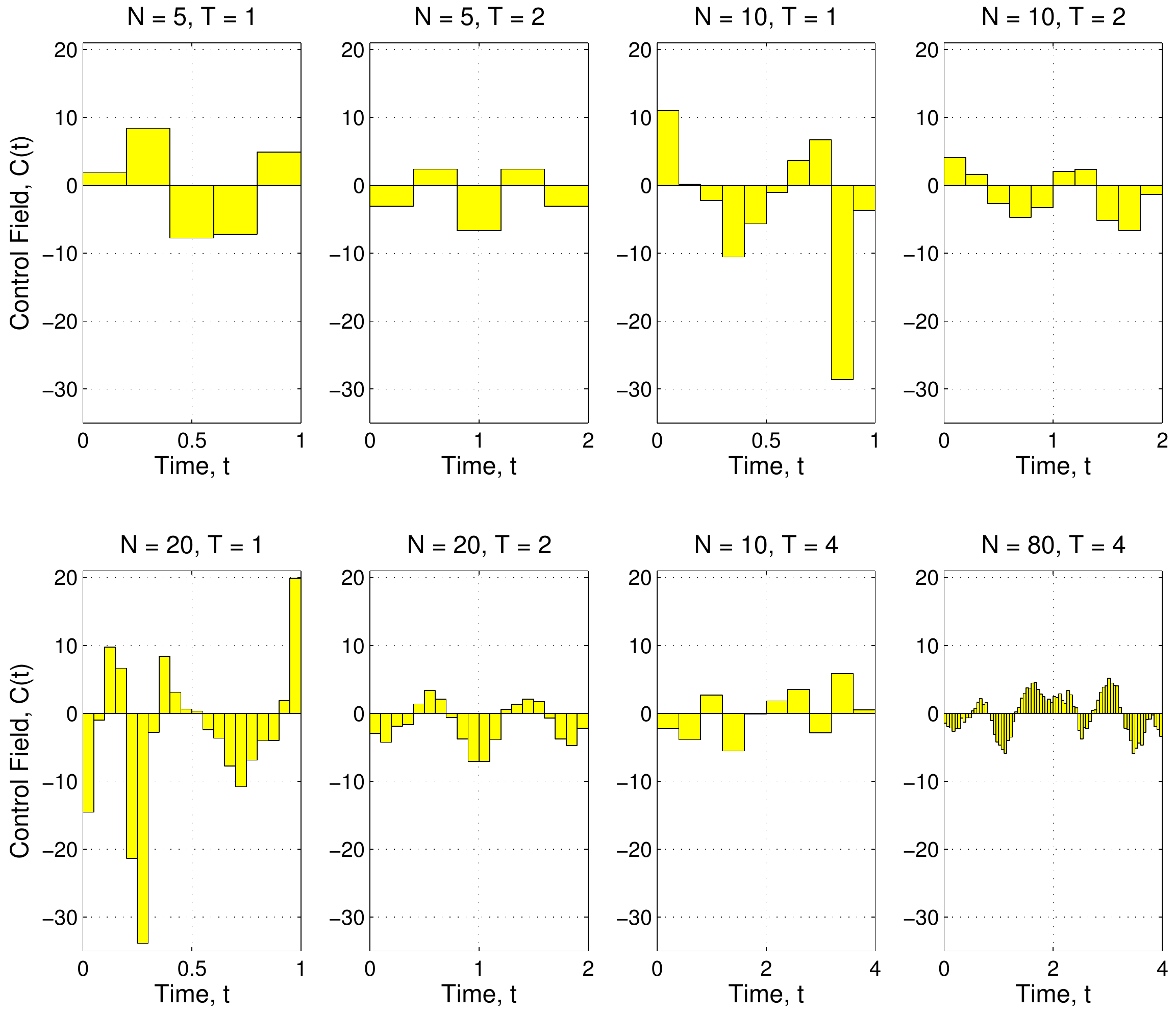}
& \hspace*{5mm}
\includegraphics[width=86.2mm]{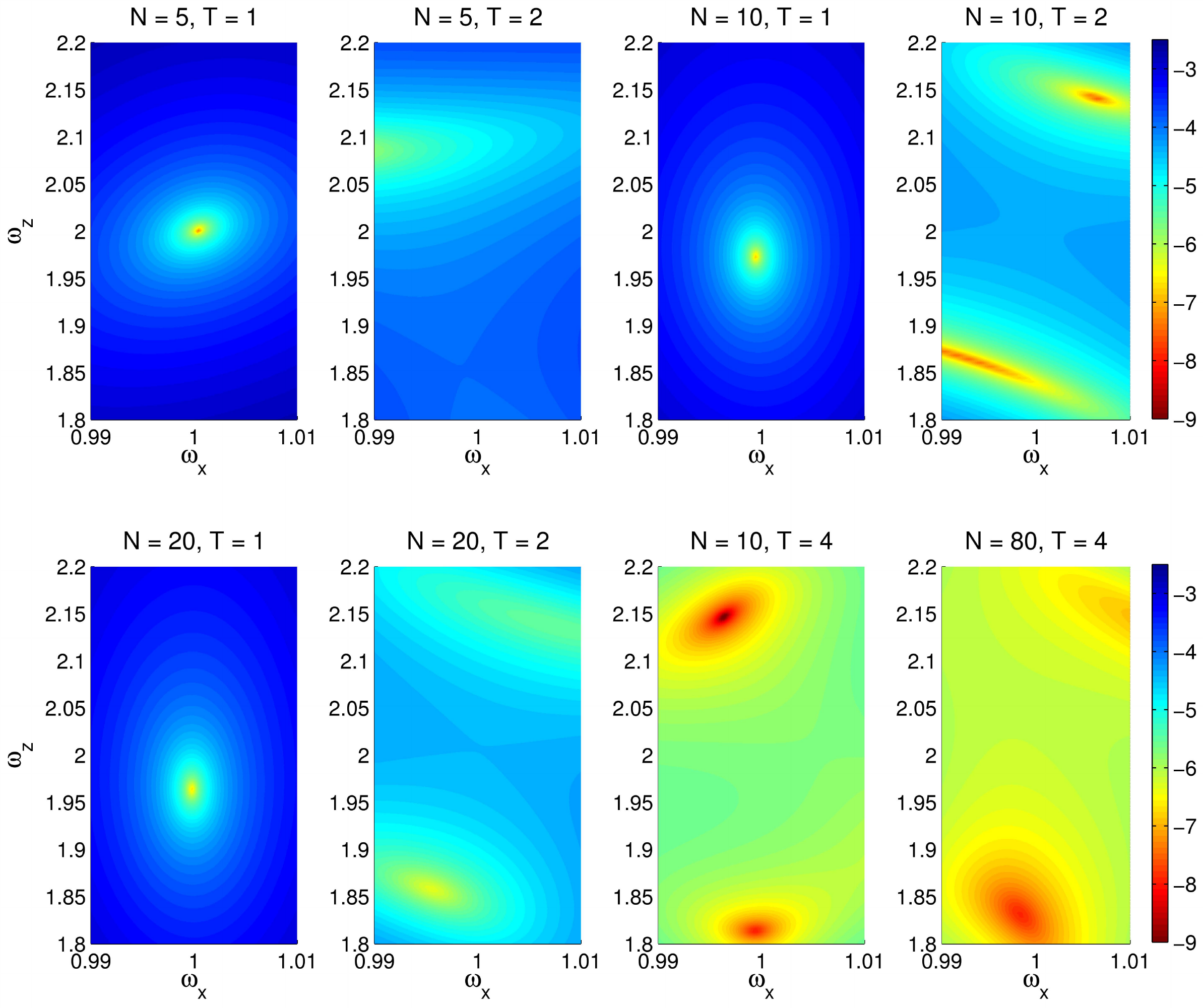}
\end{tabular}
\end{tabular}
\caption{(Color online) Left column: control fields $c(t,\theta)$ that
are solutions of the worst-case robust optimization problem
\eqref{eq:ex optxz}. Right column: logarithms of corresponding
distances, $\log_{10}\mathcal{D}(\theta,\omega)$, as functions of
$\omega_x \in [0.99, 1.01]$ and $\omega_z \in [1.8, 2.2]$. The results
are shown for eight combinations of $N$ and $T$ for each of the three
target gates of Eq.~\eqref{eq:1qu gates}: (a) the identity  gate, (b)
the Hadamard gate, and (c) the phase gate.}
\label{fig:ff}
\end{figure*}
\renewcommand{\arraystretch}{1}

Figure~\ref{fig:ff} shows control fields that are solutions of the
worst-case robust optimization problem \eqref{eq:ex optxz} for various
choices of $W$, $T$, and $N$, along with corresponding distances
$\mathcal{D}(\theta, \omega)$ that are plotted on a logarithmic scale
as functions of the parameters $\omega_x$ and $\omega_z$. Properties
of these robust controls, including logarithms of the corresponding
worst-case and average distances, the field fluence, and the maximum
field value, are listed in Table~\ref{tab:control}. For each target
gate, we present results for eight different combinations of $N$ and
$T$. With the one-qubit system of Eq.~\eqref{eq:ham1q} and the
uncertainty set $\Delta$ of Eq.~\eqref{eq:delxz}, controls with
worst-case fidelities $\mathcal{F}_{\mathrm{wc}}(\theta) \geq 0.9999$
are obtained for $N \geq 10$ and $T \geq 2$ for all three target
gates. These results demonstrate that robust, high-fidelity control is
possible with a relatively small number of control variables $N$,
provided that the final time $T$ is chosen properly.

Interestingly, the worst-case fidelity $\mathcal{F}_{\mathrm{wc}}$ can
decrease as the number of field values $N$ increases; this behavior is
seen in the results of Table~\ref{tab:control} for the identity gate
with $T = 2$ when $N$ increases from $10$ to $20$ and for the Hadamard
gate with $T = 1$ when $N$ increases from $5$ to $20$. Since the set
of controls with $N_1$ field values is a proper subset of controls
with $N_2 > N_1$ field values, these results suggest that the control
landscape for the optimization problem~\eqref{eq:ex optxz} possesses
local optima that can trap SCP searches. Thus, even more robust
solutions are, in principle, achievable by combining SCP with a
non-local algorithm capable of exploring multiple optima.

\begin{table}
\caption{\label{tab:control}Properties of control fields that are
solutions of the worst-case robust optimization problem \eqref{eq:ex
optxz}, for the three target gates of Eq.~\eqref{eq:1qu gates} and
various combinations of $N$ (the number of field values) and $T$ (the
final time).}
\begin{ruledtabular}
\begin{tabular}{rrccrr}
\multicolumn{6}{c}{Target gate: Identity} \\
$N$ & $T$ & $\log_{10}\mathcal{D}_{\mathrm{wc}}(\theta)$ &
$\log_{10}\mathcal{D}_{\mathrm{avg}}(\theta)$ & 
$\Phi(\theta)$ & $\max(\theta)$ \\
\hline
5 & 1 & -3.13 & -3.82 & 103.63 & 16.21 \\
5 & 2 & -2.35 & -3.16 & 37.24 & 7.25 \\
10 & 1 & -3.28 & -4.20 & 58.45 & 11.88 \\
10 & 2 & -5.23 & -5.79 & 51.00 & 6.59 \\
20 & 1 & -3.31 & -4.24 & 53.66 & 13.47\\
20 & 2 & -4.35 & -4.98 & 25.34 & 6.03 \\
10 & 4 & -4.62 & -5.66 & 28.96 & 4.22 \\
80 & 4 & -5.08 & -5.60 & 31.74 & 6.00 \\
\hline
\multicolumn{6}{c}{Target gate: Hadamard} \\
$N$ & $T$ & $\log_{10}\mathcal{D}_{\mathrm{wc}}(\theta)$ &
$\log_{10}\mathcal{D}_{\mathrm{avg}}(\theta)$ & 
$\Phi(\theta)$ & $\max(\theta)$ \\
\hline
5 & 1 & -2.20 & -3.08 & 32.27 & 7.59 \\
5 & 2 & -3.02 & -3.74 & 16.35 & 3.77 \\
10 & 1 & -2.17 & -3.05 & 36.93 & 8.02 \\
10 & 2 & -4.33 & -4.80 & 30.33 & 8.95 \\
20 & 1 & -2.17 & -3.06 & 34.38 & 8.64 \\
20 & 2 & -4.34 & -4.86 & 30.07 & 9.17 \\
10 & 4 & -4.06 & -4.63 & 16.60 & 2.86 \\
80 & 4 & -4.69 & -5.12 & 25.61 & 5.48 \\
\hline
\multicolumn{6}{c}{Target gate: Phase} \\
$N$ & $T$ & $\log_{10}\mathcal{D}_{\mathrm{wc}}(\theta)$ &
$\log_{10}\mathcal{D}_{\mathrm{avg}}(\theta)$ 
& $\Phi(\theta)$ & $\max(\theta)$ \\
\hline
5 & 1 & -2.77 & -3.51 & 41.98 & 8.39 \\
5 & 1 & -3.71 & -4.19 & 29.99 & 6.70 \\
10 & 1 & -2.96 & -3.55 & 116.17 & 28.62 \\
10 & 2 & -4.34 & -4.88 & 25.40 & 6.80 \\
20 & 1 & -3.02 & -3.61 & 136.06 & 33.88 \\
20 & 2 & -4.30 & -4.77 & 23.39 & 7.12 \\
10 & 4 & -5.57 & -6.02 & 46.82 & 5.87 \\
80 & 4 & -6.00 & -6.34 & 33.51 & 5.91 \\
\end{tabular}
\end{ruledtabular}
\end{table}

\section{Tradeoff between gate fidelity and control field fluence}
\label{sec:tradeoff}

The success of the optimization depends on available control
resources, and it is expected that constraints on the control field
will, generally, decrease the attainable fidelity
\cite{Moore.Rabitz.JCP.137.134113.2012, Moore.PRA.86.062309.2012}. As
a further illustration of the utility of SCP, we use it to explore the
tradeoff between the gate's worst-case fidelity and the control
field's fluence. Specifically, we consider five uncertainty sets for
$\omega$:
\begin{subequations}\label{eq:del0}
\begin{align}
& \Delta_1 = \{|\omega_x -1|\leq 0.001,\ |\omega_z -2|\leq 0.02 \}, 
\label{eq:del01} \\
& \Delta_2 = \{|\omega_x -1|\leq 0.010,\ |\omega_z -2|\leq 0.02 \}, \\
& \Delta_3 = \{|\omega_x -1|\leq 0.001,\ |\omega_z -2|\leq 0.10 \}, \\
& \Delta_4 = \{|\omega_x -1|\leq 0.010,\ |\omega_z -2|\leq 0.10 \}, \\
& \Delta_5 = \{|\omega_x -1|\leq 0.010,\ |\omega_z -2|\leq 0.20 \}.
\end{align}
\end{subequations}
These sets correspond to deterministic models with relative variations
ranging from 0.1\% to 1\% in $\omega_x$ and from 1\% to 10\% in
$\omega_z$. Note that $\Delta_5$ is the uncertainty set in
Eq.~\eqref{eq:delxz} with relative variations in $\omega_x$ and
$\omega_z$ at 1\% and 10\%, respectively. For each of the uncertainty
sets in Eqs.~\eqref{eq:del0}, we use SCP to solve for
$\theta\in\mathbb{R}^N$ from:
\begin{equation}\label{eq:scp flu}
\begin{array}{ll}
\text{maximize} 
& 
\displaystyle 
\min_{\omega \in \Delta} \left\{ \mathcal{F}(\theta, \omega)
= {\textstyle \frac{1}{4}} \left| 
\mathrm{Tr}\left( W^{\dag} U_{T} \right) \right|^{2} \right\}
\\
\text{subject to}
&
U_T\ \text{obtained from Eq.~\eqref{eq:evol-oper}},
\\
&
\Phi(\theta) \leq \gamma,
\\
&
\omega \in \Delta_m\ \text{from Eqs.~\eqref{eq:del0}}.
\end{array}
\end{equation}
The solutions of \eqref{eq:scp flu} are obtained for the target
identity gate $W_{\mathrm{I}}$, final time $T = 2$, number of field
values $N = 10$, and varying values of the fluence bound $\gamma$. For
each uncertainty set $\Delta_m$ ($m = 1,\ldots,5$), we perform a
series of SCP searches with decreasing $\gamma$. In the first SCP
search in the series, the fluence bound is set to $\gamma=\infty$
(i.e., the fluence is unconstrained) and the solution of the
optimization problem \eqref{eq:ex optxz} with the uncertainty set
\eqref{eq:delxz} is used as the initial field. In each subsequent
search in the series, $\gamma$ is set to 0.95 of the fluence of the
control field found in the previous search, and all values of the
initial field are reduced proportionally so as to match the new
fluence constraint. This process is repeated until the SCP routine
fails to achieve $\mathcal{F}_{\mathrm{wc}} \geq 0.9$ due to the
severity of the fluence constraint.

\begin{figure}[t]
\includegraphics[width=\columnwidth]{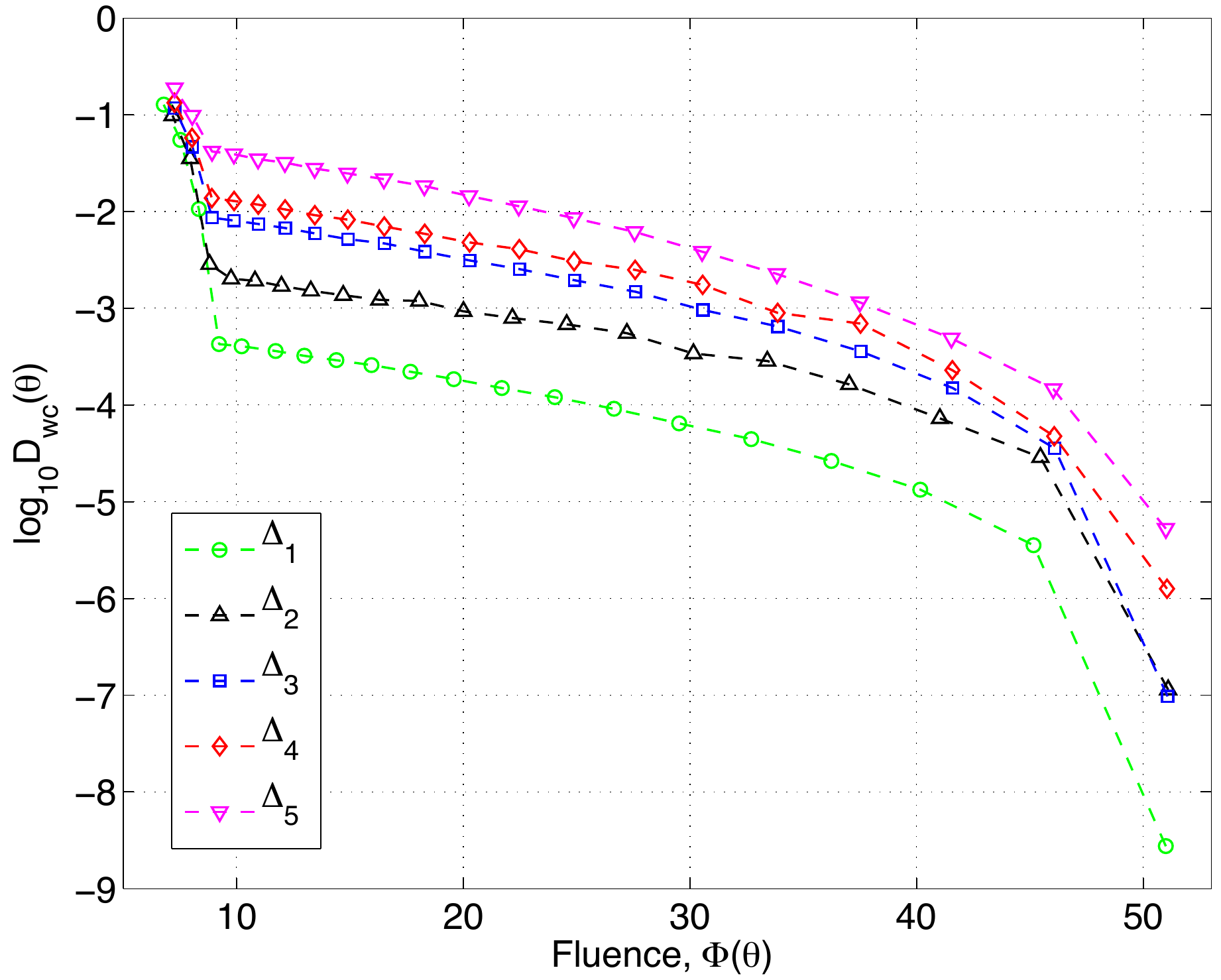}
\caption{(Color online) The logarithm of the worst-case distance,
$\log_{10}\mathcal{D}_{\mathrm{wc}}(\theta)$, versus the fluence
$\Phi(\theta)$ for control fields that are solutions of the
optimization problem \eqref{eq:scp flu}, with $W = W_{\mathrm{I}}$, $T
= 2$, and $N = 10$. Five data series, denoted by shape and color,
correspond to five uncertainty sets $\Delta_m$ of
Eqs.~\eqref{eq:del0}, as indicated in the legend.}
  \label{fig:tradeoff}
\end{figure}

Figure~\ref{fig:tradeoff} shows the resulting tradeoffs between the
logarithm of the worst-case distance,
$\log_{10}\mathcal{D}_{\mathrm{wc}}$, and the \emph{achieved} field
fluence $\Phi(\theta)$, for each of the uncertainty sets $\Delta_m$ in
\eqref{eq:del0}. The rightmost point in each series corresponds to
unconstrained fluence ($\gamma=\infty$). The rate of increase in the
distance as the fluence bound is decreased is seen to be essentially
the same for all sets $\Delta_m$. Additionally, the fluence value
where the distance abruptly changes for the worse is also about the
same: $\Phi \approx 10$. It is important to note that it is not known
if any of the tradeoff curves in Fig.~\ref{fig:tradeoff} represents a
true \emph{Pareto front} for distance versus field fluence.

The tradeoff curves in Fig.~\ref{fig:tradeoff} show that the gate
error, on average, exhibits a similar sensitivity of about one order
of magnitude to either 1\% variation in $\omega_x$ or 5\% variation in
$\omega_z$. The greater sensitivity to variations in $\omega_x$ is due
in part to the fact that in our model system \eqref{eq:ham1q}
$\omega_x$ is a direct \emph{uncertain multiplicative gain} on the
control signal. In other words, a perturbation $\tilde{\omega}_x$
around $\bar{\omega}_x = 1$ is equivalent to a perturbation
$\tilde{\omega}_x \theta$ in the control field. Following the
procedure presented in~\cite{Brif:QEC11, Moore.PRA.86.062309.2012}, a
Taylor series approximation of the fidelity up to the second order in
$\tilde{\omega}_x$ gives:
\begin{align}
\mathcal{F}(\theta,\bar{\omega}_x+\tilde{\omega}_x,\omega_z) 
& = \mathcal{F}(\theta+\tilde{\omega}_x\theta,\bar{\omega}_x,\omega_z) 
\nonumber \\
& \approx \mathcal{F}(\theta,\bar{\omega}_x,\omega_z) 
+ \tilde{\omega}_x\theta^{\mathsf{T}}\nabla_{\theta} 
\mathcal{F}(\theta,\bar{\omega}_x,\omega_z) 
\nonumber \\
& + \frac{1}{2} \tilde{\omega}_x^2 \theta^{\mathsf{T}}
\nabla_{\theta}^2\mathcal{F}(\theta,\bar{\omega}_x,\omega_z) \theta .
\label{eq:F-expansion}
\end{align}
For any control field that is a solution of the optimization problem
\eqref{eq:ex optxz} or \eqref{eq:scp flu}, the Hessian
$\nabla_{\theta}^2\mathcal{F}(\theta,\bar{\omega}_x,\omega_z)$ is
negative semi-definite and the Hessian term dominates the gradient
term. Assuming $|\tilde{\omega}_x| \leq \epsilon$, we use
Eq.~\eqref{eq:F-expansion} to obtain a lower bound on the fidelity:
\begin{align}
\mathcal{F}(\theta,\bar{\omega}_x+\tilde{\omega}_x,\omega_z) & 
\geq \mathcal{F}(\theta,\bar{\omega}_x,\omega_z) - \epsilon \left| 
\theta^{\mathsf{T}} \nabla_{\theta} 
\mathcal{F}(\theta,\bar{\omega}_x,\omega_z) \right| 
\nonumber \\
& - \frac{1}{2} \epsilon^2 \left| \theta^{\mathsf{T}} 
\nabla_{\theta}^2 \mathcal{F}(\theta,\bar{\omega}_x,\omega_z) 
\theta \right|.
\label{eq:F-bound}
\end{align} 
We evaluated the lower bound
$\underline{\mathcal{F}}(\theta,\bar{\omega}_x,\epsilon,\omega_z)$,
given by the right-hand side of Eq.~\eqref{eq:F-bound}, for control
fields whose worst-case distances are shown in
Fig.~\ref{fig:tradeoff}, using $\bar{\omega}_x = 1$, $\epsilon$ values
($\epsilon = 0.001$ and $\epsilon = 0.01$) and $\omega_z$ ranges
($\omega_z \in [1.98,2.02]$, $\omega_z \in [1.9,2.1]$, $\omega_z \in
[1.8,2.2]$) that correspond to the five uncertainty sets $\Delta_m$ of
Eqs.~\eqref{eq:del0}. Then, for each field, we minimized
$\underline{\mathcal{F}}(\theta,\bar{\omega}_x,\epsilon,\omega_z)$
over the respective $\omega_z$ range and found that the resulting
distance values coincide almost exactly with points on the
corresponding tradeoff curves in Fig.~\ref{fig:tradeoff}. This
coincidence indicates that the lower bound
$\underline{\mathcal{F}}(\theta,\bar{\omega}_x,\epsilon,\omega_z)$
approximates well the minimum of the fidelity
$\mathcal{F}(\theta,\omega)$ over the $\omega_x$ variation.

The tradeoff analysis is valuable for understanding the interplay
between constraints in control and system designs. In particular, a
limitation on the maximum field fluence can reflect not only signal
generator constraints, but also system design considerations such as
thermal loads on the system. For example, if variations in $\omega_x$
and $\omega_z$ are small, such as in the uncertainty set $\Delta_1$,
we observe from the corresponding curve in Fig.~\ref{fig:tradeoff}
that a distance value $\mathcal{D}_{\mathrm{wc}} \sim 10^{-4}$ is
possible with a fairly low fluence ($\Phi \approx 25$), and it can be
even as low as $\mathcal{D}_{\mathrm{wc}} \sim 10^{-8}$ with a higher
fluence ($\Phi\approx 50$). Attaining parameter uncertainties on the
order of $\Delta_1$ could be accomplished via material/hardware
improvements and better manufacturing/testing procedures. Certainly,
the possibility of achieving such high fidelities is a motivation to
explore these options. Thus, establishing the tradeoff between the
gate fidelity and the field fluence provides important information
about possibilities for enhancing the robust gate performance.

\section{Effect of noise}
\label{sec:noise}

When uncertainty is due to noise, performing SCP generally requires
some form of sampling from the noise distribution. If the noise is
sufficiently weak, then, based on ideas from Ref.~\cite{Brif:QEC11},
we show in Appendix~\ref{sec:wna} that an approximation can be
utilized which avoids expensive sampling. We will explore this
approach in more detail in a future paper. Here, we consider a
simplified scenario which captures some of the salient features of
robust control in the presence of noise.

Consider the Hamiltonian of Eq.~\eqref{eq:ham1q}, where the parameter
$\omega_x$ in the control term is constant, $\omega_x = 1$, and the
parameter $\omega_z$ in the drift term is a noisy time series, i.e.,
\begin{equation}\label{eq:omz-noise}
\omega_z(t) = \bar{\omega}_z+\tilde{\omega}_z(t), \ \ t \in [0,T],
\end{equation}
where $\bar{\omega}_z$ is the average value of $\omega_z(t)$ and
$\tilde{\omega}_z(t)$ is a stochastic variable obtained as the output
of a linear filter $G$ driven by stationary, Gaussian white noise
$u(t)$ with variance $\sigma^2$. Specifically, $u(t)$ satisfies:
\begin{equation}\label{eq:u-noise}
\mathbb{E}\{ u(t) \} = 0,\ \ 
\mathbb{E}\{ u(t)u(t') \} = \sigma^2 \delta(t-t'), \\
\end{equation}
the filter action is
\begin{equation}\label{eq:domz}
\tilde{\omega}_z(t)=(G \ast u)(t), \ \ t\in(-\infty,T], \\
\end{equation}
$G$ is a linear first-order filter with the transfer function:
\begin{equation}\label{eq:filter}
G(s) = 1/(s\tau+1) , 
\end{equation}
and $\tau$ is the filter time-constant.

The average gate fidelity $\mathcal{F}_{\mathrm{avg}}(\theta) =
\mathbb{E}\{ \mathcal{F}(\theta,\omega_z) \}$ under a noise process
affecting $\omega_z$ can be evaluated using random sampling from the
noise distribution:
\begin{equation}\label{eq:Favg-noise-MC}
\mathcal{F}_{\mathrm{avg}}(\theta) \approx 
\sum_{l=1}^L \mathcal{F}(\theta,\omega_z^{(l)}) .  
\end{equation} 
Here, $\omega_z \in \mathbb{R}^M$ is the vector whose elements
represent a piecewise-constant approximation of the time series
$\omega_z(t)$ with a uniform time step $\tilde{h} = T/M$,
$\omega_z^{(l)} = \bar{\omega}_z + \tilde{\omega}_z^{(l)}$ is the
vector corresponding to the $l$th realization of the noise process,
and $L$ is the number of noise realizations in the sample. Another
method for evaluating $\mathcal{F}_{\mathrm{avg}}(\theta)$ is the weak
noise approximation described in Appendix~\ref{sec:wna}. Specifically,
using the Taylor series expansion of the fidelity about
$\bar{\omega}_z$ up to the second order in $\tilde{\omega}_z$ and
assuming that $\tilde{\omega}_z$ has zero mean and covariance matrix
$C = \mathbb{E}\{\tilde{\omega}_z \tilde{\omega}_z^{\mathsf{T}}\}$, we
obtain [cf.~Eq.~\eqref{eq:fav approx}]:
\begin{equation}\label{eq:Favg-noise-WNA}
\mathcal{F}_{\mathrm{avg}}(\theta) \approx 
\mathcal{F}(\theta,\bar{\omega}_z) - 
\frac{1}{2} \mathrm{Tr}(C R_{\omega_z\omega_z}),  
\end{equation}  
where $R_{\omega_z \omega_z} = -\nabla_{\omega_z}^2
\mathcal{F}(\theta,\bar{\omega}_z)$ is the negative Hessian
matrix. For the filtered noise process of
Eqs.~\eqref{eq:u-noise}--\eqref{eq:filter}, elements of the covariance
matrix $C$ are given by:
\begin{equation}\label{eq:cov}
C_{m m'} = \tilde{\sigma}^2 
\frac{1-\alpha}{1+\alpha} \alpha^{|m-m'|},\ \
m,m' = 1,\ldots,M,
\end{equation}
where $\tilde{\sigma}^2 = \sigma^2/\tilde{h}$ and $\alpha =
\exp(-\tilde{h}/\tau)$.

\begin{figure}[htbp]
\includegraphics[width=\columnwidth]{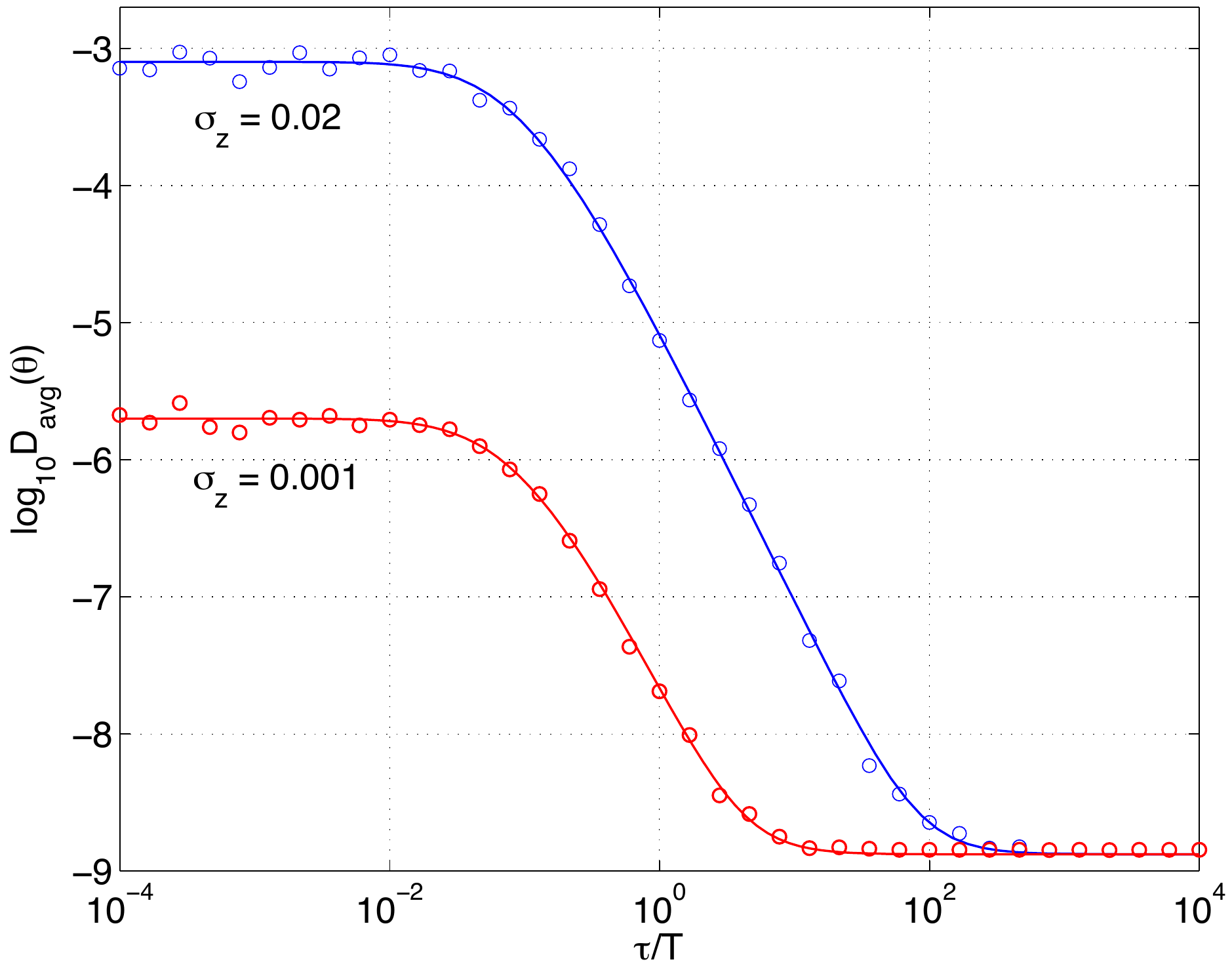}
\caption{(Color online) The logarithm of the average distance,
$\log_{10}\mathcal{D}_{\mathrm{avg}}(\theta)$, versus $\tau/T$, where
$\tau$ is the noise filter time-constant from Eq.~\eqref{eq:filter}.
The control field used here is a solution of the optimization problem
\eqref{eq:scp flu} with $W = W_{\mathrm{I}}$, $T = 2$, $N = 10$,
$\Delta_1$ of Eq.~\eqref{eq:del01}, and no fluence constraint
($\gamma=\infty$). The distance is averaged under the noise process
described by Eqs.~\eqref{eq:omz-noise}--\eqref{eq:filter} with
$\bar{\omega}_z = 2$ and two $\sigma$ values: $\sigma = 0.001$ (red
line and circles) and $\sigma = 0.02$ (blue line and circles). The
weak noise approximation (solid lines) computed using
Eqs.~\eqref{eq:Favg-noise-WNA} and \eqref{eq:cov} is in a very good
agreement with simulated data (circles) obtained using random sampling
from the noise distribution according to
Eq.~\eqref{eq:Favg-noise-MC}.}
  \label{fig:noise}
\end{figure}

It is interesting to analyze how a control field designed to be robust
for a deterministic uncertainty model performs in the presence of
noise. For example, consider the control field that is a solution of
the optimization problem \eqref{eq:scp flu} with $W = W_{\mathrm{I}}$,
$T = 2$, $N = 10$, $\Delta_1$ of Eq.~\eqref{eq:del01}, and no fluence
constraint ($\gamma=\infty$); this field corresponds to the rightmost
point on the bottom curve in Fig.~\ref{fig:tradeoff}. For this field,
we use both the random sampling method of Eq.~\eqref{eq:Favg-noise-MC}
and weak noise approximation of Eq.~\eqref{eq:Favg-noise-WNA} to
evaluate the average fidelity $\mathcal{F}_{\mathrm{avg}}(\theta)$
under the noise process of Eqs.~\eqref{eq:u-noise}--\eqref{eq:filter}
with $\bar{\omega}_z = 2$, $\sigma \in \{0.001, 0.02\}$, and various
values of $\tau$. Figure~\ref{fig:noise} shows the corresponding
values of $\log_{10}\mathcal{D}_{\mathrm{avg}}(\theta)$, for a range
of filter time-constants relative to the control time, $\tau/T \in
[10^{-4},10^{4}]$. We observe an excellent agreement between the weak
noise approximation (solid lines) and simulated data from random
sampling (circles). 

Equations~\eqref{eq:Favg-noise-WNA} and \eqref{eq:cov} can be further
used to investigate the asymptotic behavior in the limits of
low-bandwidth and high-bandwidth filter. For $\tau/T \gg 1$, the
filter bandwidth is very low, and the noise is effectively blocked. In
this limit, all elements of $C$ are approximately zero, and
$\mathcal{D}_{\mathrm{avg}}(\theta) \approx
\mathcal{D}(\theta,\bar{\omega}_z) \approx 10^{-8.88}$ is independent
of $\sigma$. For $\tau/T \ll 1$, the filter bandwidth is very high,
which allows for the white noise to pass through unaltered. In this
limit, $C$ is proportional to the identity matrix: $C_{m m'} \approx
\tilde{\sigma}^2 \delta_{m m'}$, and
$\mathcal{F}_{\mathrm{avg}}(\theta) \approx
\mathcal{F}(\theta,\bar{\omega}_z) - \frac{1}{2} \tilde{\sigma}^2
\mathrm{Tr} (R_{\omega_z\omega_z})$. For a control field
$c(t,\theta^{\star})$ which is globally optimal for the objective of
maximizing $\mathcal{F}(\theta,\bar{\omega}_z)$, each diagonal matrix
element of $R_{\omega_z\omega_z}$ equals $2 \tilde{h}^2$, and we
obtain a simple analytic result:
\begin{equation}\label{eq:wn-limit}
\mathcal{D}_{\mathrm{avg}}(\theta^{\star}) \approx \sigma^2 T .  
\end{equation}
The field that we use here is not exactly $\theta^{\star}$, but the
value $\mathcal{D}(\theta,\bar{\omega}_z) \approx 10^{-8.88}$ is
sufficiently close to the optimum for the result of
Eq.~\eqref{eq:wn-limit} to be a very good approximation. Then, with $T
= 2$, we obtain $\mathcal{D}_{\mathrm{avg}} \approx 2 \times 10^{-6}
\approx 10^{-5.70}$ for $\sigma = 0.001$ and
$\mathcal{D}_{\mathrm{avg}} \approx 8 \times 10^{-4} \approx
10^{-3.10}$ for $\sigma = 0.02$. The asymptotic results in both limits
are very well confirmed by the data shown in Fig.~\ref{fig:noise}. 

The results of Fig.~\ref{fig:noise} also show that the control field,
though not designed for stochastic uncertainty, nonetheless performs
admirably. In fact, we used SCP to find control fields that are
specifically robust against noise in $\omega_z$, but did not obtain a
significant improvement. While in this example a robust control
designed for a deterministic uncertainty model also works well against
a stationary stochastic process, we do not know whether this behavior
holds in general.

\section{Summary}
 
Using SCP we demonstrated that it is possible to generate
high-fidelity quantum gates with a substantial robustness against
uncertainties, while simultaneously using limited control resources
such as field amplitude, bandwidth, and fluence. Designing such robust
control fields requires a specific knowledge of the range and
character of the uncertainties, a process referred to in the control
theory literature as ``uncertainty modeling.'' Although we focused on
a one-qubit system, even this simple example clearly shows the strong
effect of control constraints on the attainable degree of
robustness. Our analysis of this system also revealed that a control
field designed for a deterministic (set-membership) uncertainty model
can be quite effective against stochastic uncertainty (noise).  

This work shows that SCP is useful for exploring possible improvements
in the robust gate performance for different values and ranges
available for both control and system designs. Specifically, SCP makes
it possible to quantify a variety of tradeoffs between constraints on
control and system parameters. For example, one can determine how many
control variables are required to achieve a desired worst-case
fidelity for a given uncertainty range or, alternatively, how tight
should be the uncertainty range for a given limitation on the maximum
field fluence. Such tradeoff analysis could reveal a combination of
physical design and robust control design resulting in a ``sweet
spot'' amongst the possibilities.

Of course, SCP is not the only approach to finding locally optimal
solutions to non-convex problems. An important advantage of SCP is the
ease by which various uncertainty models and constraints on design
variables can be directly incorporated in the local convex
optimization step of the algorithm. It would be desirable to develop a
hybrid approach, integrating SCP with a non-local optimization method,
in order to make it possible to search among multiple solutions.

The array of results presented here hopefully herald what would be
seen in more complex systems, involving multiple qubits, controlled
ancillae, coupling to a bath, and so on. In addition, the results also
begin to provide an insight into unanticipated control field
structures. Many of these potentialities are under consideration at
present and will be forthcoming.

\acknowledgments 

We gratefully acknowledge helpful discussions with Kevin Young
(SNL-CA), Kaveh Khodjasteh, and Lorenza Viola (Dartmouth College).
This work was supported by the Laboratory Directed Research and
Development program at Sandia National Laboratories. Sandia is a
multi-program laboratory managed and operated by Sandia Corporation, a
wholly owned subsidiary of Lockheed Martin Corporation, for the United
States Department of Energy's National Nuclear Security Administration
under contract DE-AC04-94AL85000. RLK acknowledges support from the
ARO MURI Grant W911NF-11-1-0268 to USC and the Intelligence Advanced
Research Projects Activity (IARPA) via Department of Interior National
Business Center contract number D11PC20165. The U.S. Government is
authorized to reproduce and distribute reprints for Governmental
purposes notwithstanding any copyright annotation thereon. Disclaimer:
The views and conclusions contained herein are those of the authors
and should not be interpreted as necessarily representing the official
policies or endorsements, either expressed or implied, of IARPA,
DoI/NBC, or the U.S. Government.

\appendix

\section{Convex optimization}
\label{sec:scpcode}

The convex optimization step in the SCP algorithm can be equivalently
expressed as:
\begin{equation}\label{eq:cvx grad}
\begin{array}{ll}
\text{maximize} 
& 
f_0
\\
\text{subject to}
& f_i + \tilde{\theta}^{\mathsf{T}} g_i \geq f_0,
\ \ i=1,\ldots,L,
\\
& \theta + \tilde{\theta}\in\Theta,\ 
\tilde{\theta} \in \tilde{\Theta}_{\mathrm{trust}} ,
\end{array}
\end{equation}
where $f_i = \mathcal{F}(\theta,\delta_i)$ and $g_i = \nabla_{\theta}
\mathcal{F}(\theta,\delta_i)$. Both $\tilde{\theta}$ and $f_0$ are now
the optimization variables. If both $\Theta$ and
$\tilde{\Theta}_{\mathrm{trust}}$ bound their respective elements in a
``box'' in $\mathbb{R}^N$, then \eqref{eq:cvx grad} is a \emph{linear
program}.

The Hessian can be employed in the optimization step by using its
negative definite part, $R_i = -[\nabla^2_{\theta}
\mathcal{F}(\theta,\delta_i)]_{-}$ where $[\cdot]_{-}$ retains only
the negative eigenvalues of the Hessian, specifically, $R_i =
V_iV_i^{\mathsf{T}}$ with $V_i \in \mathbb{R}^{N\times r}$ where $r$
is the number of strictly negative eigenvalues of the Hessian less
than or equal to a chosen threshold. Then the worst-case fidelity
constraint can be formulated as:
\begin{equation}\label{eq:hess-constraint}
f_i + \tilde{\theta}^{\mathsf{T}} g_i - {\textstyle \frac{1}{2}} 
\tilde{\theta}^{\mathsf{T}} V_iV_i^{\mathsf{T}} \tilde{\theta} 
\geq f_0,\ \ i=1,\ldots,L.
\end{equation}
Each of the inequalities in Eq.~\eqref{eq:hess-constraint} is
equivalent to a linear-matrix-inequality in the variables
$\{\tilde{\theta},f_0\}$ \cite{BoydV:04}: $Q_i(\tilde{\theta},f_0)
\geq 0$, where
\begin{equation}\label{eq:Q-matrix}
Q_i(\tilde{\theta},f_0) = 
\begin{bmatrix}
f_i - f_0 - \tilde{\theta}^{\mathsf{T}} g_i &
\tilde{\theta}^{\mathsf{T}}V_i/\sqrt{2}
\\
V_i^{\mathsf{T}}\tilde{\theta}/\sqrt{2} & I_r
\end{bmatrix},
\end{equation}
and $I_r$ is the $r \times r$ identity matrix. The optimization step
in SCP is now given by the \emph{semidefinite program}:
\begin{equation}\label{eq:cvx hess}
\begin{array}{ll}
\text{maximize} & f_0
\\
\text{subject to}
&
Q_i(\tilde{\theta},f_0) \geq 0, \ \ i=1,\ldots,L,
\\
&
\theta + \tilde{\theta} \in \Theta,\ 
\tilde{\theta} \in \tilde{\Theta}_{\mathrm{trust}}.
\end{array}
\end{equation}
The optimization problems \eqref{eq:cvx grad} and \eqref{eq:cvx hess}
are now expressed in standard forms suitable for use with existing
software specially developed for these classes of convex
optimization. In particular, YALMIP \cite{yalmip:04} and CVX
\cite{CVXurl,CVXpaper} are convex compilers compatible with
MATLAB. Using these software tools makes it very easy to code the
convex optimization problems almost exactly as expressed
mathematically. These compilers call convex solvers such as SDPT-3
\cite{SDPT3} and SeDuMi \cite{SEDUMI} which have been developed and in
use for many years, and as a result are generally efficient and
reliable. There are limits imposed by both memory and speed for a
particular problem instance and computer platform. In these cases it
could be necessary to use or develop specialized versions with
modifications that take into account the specific underlying structure
of the problem.

\section{Signal generation}
\label{sec:signal}

In general, the control field $c(t,\theta)$ is the output of a signal
generation device. As an example, consider a field generated by a
device with rate $\nu$ and piecewise-constant commands $\theta$:
\begin{subequations}\label{eq:csig}
\begin{align}
& \dot{c}(t,\theta) = \nu [\bar{c}(t,\theta)-c(t,\theta)], \ \ 
c(0) = 0 , \\
& \bar{c}(t,\theta) = \theta_{k} \ \ \text{for} \ \ 
t \in (t_{k-1},t_k] , \ \ k=1,\ldots,N,
\end{align}
\end{subequations}
where $t_k = k h$ and $h = T/N$. The field in this example can be
expressed in a general form:
\begin{equation}\label{eq:clindyn}
c(t,\theta) = \sum_{k=1}^N s_k(t) \theta_k
= s(t)^{\mathsf{T}} \theta,\ \ t \in [0,T] .
\end{equation}
This expression holds for \emph{any} signal generation well
represented by known linear dynamics whose input is a finite sequence
of control commands $\{\theta_k\}$ at a uniform sampling rate. The
linear dynamics are captured in the shape-function vector $s(t) \in
\mathbb{R}^N$. For example, for the field of Eqs.~\eqref{eq:csig}, the
elements of $s(t)$ are given by
\begin{equation}\label{eq:s-csig}
s_k(t) = \left\{ \begin{array}{l}
0 \ \ \text{for} \ \ t \leq t_k, \\
1 - e^{-\nu (t-t_{k-1})} \ \ \text{for} \ \ t \in (t_{k-1},t_k], \\
(1-e^{-\nu h})e^{-\nu (t-t_k)} \ \ \text{for} \ \ t > t_k .
\end{array} \right.
\end{equation} 
In the limit of very fast dynamics ($\nu\to\infty$), the element
$s_k(t)$ of Eq.~\eqref{eq:s-csig} becomes the indicator function of
the interval $(t_{k-1},t_k]$, and the control field $c(t,\theta)$ is
piecewise-constant over $N$ uniform time intervals of width $h$, as
given by Eq.~\eqref{eq:cpwc}. 

Generally, when the dynamics of the signal generation device have an
appreciable effect on the shapes of $\{ s_k(t) \}$, the numerical
integration of the Schr\"{o}dinger equation~\eqref{eq:schro} would
require using a time step over which the field $c(t,\theta)$ does not
change much, i.e., finer than the command interval $h$.

Note that for any field of the form \eqref{eq:clindyn}, the control
constraint sets in Table~\ref{tab:thfeas} are convex. For example, the
constraint on the field fluence can be expressed as $\Phi(\theta) =
\theta^{\mathsf{T}} B \theta\leq\gamma$ with $B = \int_0^T
s(t)s(t)^{\mathsf{T}} d t$.

The field form of Eq.~\eqref{eq:clindyn} can be further generalized by
considering multiple command vectors $\{ \theta_i \}$ and
shape-function vectors $\{ s_i(t) \}$, i.e.,
\begin{equation}\label{eq:clcd-th}
c(t,\theta) = \sum_{i=1}^K s_i(t)^{\mathsf{T}} \theta_i .
\end{equation}
For example, laser pulse shaping in a liquid crystal modulator
generates a control field of the form: 
\begin{equation}\label{eq:clcd} 
c(t) = A_0(t) \sum_{i=1}^K a_i \sin(\omega_i t + \phi_i) , 
\end{equation}
where the envelope function $A_0(t)$ and frequencies $\{\omega_i\}$
are fixed, while amplitudes $\{a_i\}$ and phases $\{\phi_i\}$ are the
control variables. The field~\eqref{eq:clcd} can be equivalently
expressed in the form~\eqref{eq:clcd-th}, where $s_i(t) = A_0(t)
[\sin(\omega_it),\cos(\omega_it)]^{\mathsf{T}}$ and $\theta_i = a_i
[\cos\phi_i,\sin\phi_i]^{\mathsf{T}}$. 

For a control field of the form~\eqref{eq:clcd}, constraints are
typically imposed on the amplitudes $\{a_i\}$ and, since
$\|\theta_i\|_2 = a_i$, they can be equivalently expressed as
constraints on $\theta$.  For example, the magnitude constraint $a_i
\leq a_{\max}$ is equivalent to the convex set $\|\theta_i\|_2 \leq
a_{\max}$. However, the constraint that all the amplitudes are the
same, i.e., $a_i = a_0$ is equivalent to the non-convex set
$\|\theta_i\|_2 = a_0$. This problem can be circumvented by using the
constraint set $\|\theta_i\|_2 \leq a_0$ which is a \emph{convex
relaxation} \cite{BoydV:04} of the actual non-convex one; then SCP
will return a local solution to the relaxed problem. Some relaxations
can be proven to be optimal, but that is not known here and hence a
post-optimization analysis is required.

\section{Weak noise approximation}
\label{sec:wna}

When the noise variance is small, it is possible to avoid the
expensive simulation of noise realizations drawn from the underlying
distribution. The approach for evaluating the effect of weak noise,
the basics of which are presented here, was introduced in
Refs.~\cite{Brif:QEC11, Moore.PRA.86.062309.2012} and will be explored
in depth in a subsequent paper.

Consider an $n$-level quantum system with the Hamiltonian: 
\begin{equation}\label{eq:hamcw}
H(t) = c(t) H_c + w(t) H_w,
\end{equation}
where $c(t)$ and $w(t)$ are, respectively, the control field and the
noisy field (real-valued functions of time defined on the interval
$[0,T]$). We assume that $c(t) = c(t,\theta)$ is a piecewise-constant
function of the form~\eqref{eq:cpwc}. Let the elements $\{ w_m \}$ of
the vector $w\in\mathbb{R}^M$ represent a piecewise-constant
approximation of $w(t)$, i.e.,
\begin{equation}\label{eq:wunc}
w(t) = w_m \ \ \text{for}\ \ 
t \in (\tilde{t}_{m-1}, \tilde{t}_m] ,\ \ 
m = 1,\ldots,M,
\end{equation} 
where $\tilde{t}_m = m \tilde{h}$ and $\tilde{h} = T/M$. Assuming that
$M \geq N$ and $p = M/N$ is an integer, the control can be represented
as:
\begin{equation}\label{eq:c-vec}
c(t) = c_m \ \ \text{for}\ \ 
t \in (\tilde{t}_{m-1}, \tilde{t}_m] ,\ \ 
m = 1,\ldots,M,
\end{equation} 
where $\{c_m\}$ are the elements of the vector $c = \theta \otimes e_p
\in \mathbb{R}^M$ and $e_p$ denotes the vector of ones of length
$p$. Analogous to Eqs.~\eqref{eq:evol-oper} and
\eqref{eq:evol-prop-one-step}, the time-evolution operator is given by
\begin{align}
& U(\tilde{t}_m) = U(\tilde{t}_m,\tilde{t}_{m-1}) \cdots 
U(\tilde{t}_2,\tilde{t}_1) U(\tilde{t}_1,\tilde{t}_0) , 
\label{eq:evol-oper-2} \\
& U(\tilde{t}_m,\tilde{t}_{m-1}) = \exp \left[ 
-i \tilde{h} (c_m H_c + w_m H_w) \right],
\label{eq:evol-prop-one-step-2}
\end{align}
and, in particular, $U_T = U(\tilde{t}_M)$. The gate fidelity is
\begin{equation}\label{eq:fid cw}
\mathcal{F}(\theta,w) = \frac{1}{n^2} \left|\mathrm{Tr}\left(
W^{\dag} U_T \right)\right|^2 .
\end{equation}

Assume that the noisy field has the form $w = \bar{w} + \tilde{w}$,
where $\bar{w} \in \mathbb{R}^M$ is a deterministic mean and
$\tilde{w} \in \mathbb{R}^M$ is a stochastic variable that represents
a stationary noise process. For a specified control $\theta$, the
Taylor series expansion of the fidelity about $\bar{w}$ up to the
second order in $\tilde{w}$, gives the approximation:
\begin{equation}\label{eq:fdw}
\mathcal{F}(\theta,w) \approx \mathcal{F}(\theta,\bar{w}) 
+ \tilde{w}^{\mathsf{T}} g_w 
- \frac{1}{2}\tilde{w}^{\mathsf{T}}R_{ww}\tilde{w} ,
\end{equation}
where $g_w = \nabla_w\mathcal{F}(\theta,\bar{w}) \in \mathbb{R}^M$ is
the gradient vector and $R_{ww} =
-\nabla^2_w\mathcal{F}(\theta,\bar{w}) \in \mathbb{R}^{M \times M}$ is
the \emph{negative} Hessian matrix. Assume that the stochastic
variable $\tilde{w}$ has zero mean and covariance matrix $C \in
\mathbb{R}^{M \times M}$, i.e.,
\begin{equation}\label{eq:dw}
\mathbb{E}\{\tilde{w}\} = 0,\ \ \
\mathbb{E}\{\tilde{w} \tilde{w}^{\mathsf{T}}\} = C.
\end{equation}
The fidelity averaged over all noise realizations is given by the
statistical expectation: $\mathcal{F}_{\mathrm{avg}}(\theta) =
\mathbb{E}\{ \mathcal{F}(\theta,w) \}$. Using Eqs.~\eqref{eq:fdw} and
\eqref{eq:dw}, we obtain the weak noise approximation for the average
fidelity:
\begin{equation}\label{eq:fav approx}
\mathcal{F}_{\mathrm{avg}}(\theta) \approx 
\mathcal{F}(\theta,\bar{w}) - \frac{1}{2} \mathrm{Tr}(C R_{ww}).
\end{equation}
Since the dependence on noise in Eq.~\eqref{eq:fav approx} is only
through the covariance matrix, the evaluation of
$\mathcal{F}_{\mathrm{avg}}(\theta)$ via this approximation does not
require random sampling from the noise distribution,
providing a huge advantage in numerical efficiency.

For Gaussian white noise with variance $\sigma^2$, the covariance
matrix is given by $C = (\sigma^2/\tilde{h}) I_M$, and
Eq.~\eqref{eq:fav approx} yields:
\begin{equation}\label{eq:fav approx wn}
\mathcal{F}_{\mathrm{avg}}^{(\mathrm{wn})}(\theta) \approx 
\mathcal{F}(\theta,\bar{w}) - 
\frac{\sigma^2}{2 \tilde{h}} \mathrm{Tr}(R_{ww}).
\end{equation}
For a control $\theta^{\star}$ which is globally optimal for the
objective of maximizing $\mathcal{F}(\theta,\bar{w})$, all diagonal
matrix element of $R_{ww}$ are equal to each other:
\begin{equation}\label{eq:R-opt}
(R_{ww})_{m m} = \frac{2 \tilde{h}^2}{n} \mathrm{Tr}( H_w^2 )
- \frac{2 \tilde{h}^2}{n^2} [\mathrm{Tr}(H_w )]^2 ,
\ \ \forall m.
\end{equation}
If the operator $H_w$ is traceless, substituting Eq.~\eqref{eq:R-opt}
into Eq.~\eqref{eq:fav approx wn} leads to a simple analytical
expression: 
\begin{equation}\label{eq:fav approx wn opt}
\mathcal{F}_{\mathrm{avg}}^{(\mathrm{wn})}(\theta^{\star}) \approx 
1 - \frac{1}{n} \mathrm{Tr}( H_w^2 ) \sigma^2 T .
\end{equation}
This result shows that robustness against additive white noise can be
improved only by reducing the control duration $T$; however, this can
be done only as long as $T \geq T^{\ast}$, where $T^{\ast}$ is a
critical value below which the nominal objective is not reachable
\cite{Moore.PRA.86.062309.2012}. In a quantum information system,
$H_w$ is typically given by a tensor product of Pauli matrices and
identity operators for individual qubits. In this case, $H_w^2 = I_n$,
and Eq.~\eqref{eq:fav approx wn opt} is further simplified:
\begin{equation}\label{eq:fav approx wn opt 2}
\mathcal{F}_{\mathrm{avg}}^{(\mathrm{wn})}(\theta^{\star}) \approx 
1 - \sigma^2 T .
\end{equation}

The weak noise approximation together with a similar expansion for a
small control change (from $\theta$ to $\theta+\tilde{\theta}$) can be
used in the optimization step of SCP for designing controls robust to
a stochastic uncertainty model. Expanding the fidelity about
$\{\theta,\bar{w}\}$ up to the second order in
$\{\tilde{\theta},\tilde{w}\}$ gives:
\begin{equation}
\mathcal{F}(\theta+\tilde{\theta},w) \approx f 
+ \tilde{x}^{\mathsf{T}} g
- \frac{1}{2} \tilde{x}^{\mathsf{T}} R \tilde{x} , 
\label{eq:fdthdw} 
\end{equation}
where
\begin{equation}
\tilde{x} = \begin{bmatrix} 
\tilde{\theta} \\ \tilde{w} \end{bmatrix} , \ \
g = \begin{bmatrix} g_{\theta} \\ g_{w} \end{bmatrix} , \ \
R = \begin{bmatrix} R_{\theta\theta} & R_{\theta w} \\ 
R_{w \theta} & R_{ww} \end{bmatrix} ,
\end{equation} 
$f = \mathcal{F}(\theta,\bar{w})$ is the fidelity, $g_a =
\nabla_a\mathcal{F}(\theta,\bar{w})$ are gradient vectors, and $R_{a
b} = -\nabla_a \nabla_b \mathcal{F}(\theta,\bar{w})$  are negative
Hessian matrices ($a,b \in \{\theta,w\}$). Given a model of the noise
distribution, we can then pose the robust optimization problem:
\begin{equation}\label{eq:rbst stoch}
\begin{array}{ll}
\text{maximize} & \gamma \\
\text{subject to} &
\mathrm{Prob}\{ \mathcal{F}(\theta+\tilde{\theta},w) \geq \gamma \} 
\geq \eta,\ \
\tilde{\theta}\in\Theta .
\end{array}
\end{equation}
Assume further that the stochastic variable $\tilde{w}$ has a
zero-mean Gaussian distribution with covariance matrix $C$, i.e.,
satisfies Eq.~\eqref{eq:dw}, with $\| C \| = O(\sigma^2)$. Following
the approach to robust optimization described in
\cite[Ch.4]{BoydV:04}, the problem \eqref{eq:rbst stoch} is
equivalent, up to $O(\sigma^2)$, to the \emph{second order cone
program} (SOCP) with optimization variables $\tilde{\theta}$ and
$\gamma$:
\begin{equation}\label{eq:socp}
\begin{array}{ll}
\text{maximize} & \gamma \\
\text{subject to} &
\bar{\mathcal{F}}(\theta+\tilde{\theta}) \geq \gamma 
+ \Phi^{-1}(\eta) V^{1/2},\ \
\tilde{\theta}\in\Theta ,
\end{array}
\end{equation}
where
\begin{align}
& \bar{\mathcal{F}}(\theta+\tilde{\theta}) = 
f + \tilde{\theta}^{\mathsf{T}} g_{\theta} 
- \frac{1}{2} \tilde{\theta}^{\mathsf{T}} 
R_{\theta\theta} \tilde{\theta}
- \frac{1}{2}\mathrm{Tr}(CR_{ww}), \\
& V = (R_{w\theta}\tilde{\theta} - g_w)^{\mathsf{T}} 
C (R_{w\theta}\tilde{\theta} - g_w),
\end{align}
and $\Phi(\eta)$ is the cumulative distribution function for the
normal Gaussian density. The SOCP of Eq.~\eqref{eq:socp} can be used
in the optimization step in Algorithm~\ref{alg:scp}. The use of the
weak noise approximation makes this approach very numerically
efficient. Indeed, calculations at each new control $\theta$ require
only the knowledge of the noise covariance matrix, thus eliminating
the need for random sampling from the noise distribution. A full
exploration of this approach will be forthcoming.


\bibliographystyle{apsrev4-1}
\bibliography{scp}

\end{document}